\documentclass[11pt]{article}
\usepackage{jheppub}
\usepackage{bm,bbm}
\usepackage{booktabs,amsmath}
\usepackage{accents}
\usepackage{multirow}
\usepackage{graphicx}
\usepackage{braket}
\usepackage{mathtools}
\usepackage{comment}
\usepackage{autobreak}
\usepackage{subcaption}
\usepackage{enumerate}
\usepackage{blkarray}
\usepackage{longtable}
\usepackage{adjustbox}
\usepackage{makecell}

\usepackage[dvipsnames]{xcolor}

\definecolor{defectblue}{RGB}{199,224,231}
\definecolor{bulkgray}{RGB}{245,245,245}
\definecolor{interyellow}{RGB}{246,234,200}

\usepackage{tikz}
\usetikzlibrary{patterns}
\usetikzlibrary{arrows,shapes,positioning}
\usetikzlibrary{decorations.markings}
\usetikzlibrary{decorations.pathmorphing}
\usetikzlibrary{calc}
\tikzset{snake it/.style={decorate, decoration=snake}}

\hypersetup{
    colorlinks=true,
    linkcolor=blue,
    filecolor=magenta,      
    urlcolor=cyan,
    pdfpagemode=FullScreen,
    }

\usepackage{arydshln}
\usepackage{mathtools}

\edef\restoreparindent{\parindent=\the\parindent\relax}
\usepackage{parskip}
\restoreparindent
\usepackage{ascmac}
\usepackage{mathrsfs}

\usepackage{amsthm}
\newtheoremstyle{break}
  {\topsep}{\topsep}%
  {\upshape}{}%
  {\bfseries}{}%
  {\newline}{}%
\theoremstyle{break}

\def\Tr{{\rm Tr}}

\def\i{{\rm i}}
\def\CA{{\cal A}}
\def\CB{{\cal B}}

\def\CD{{\cal D}}
\def\CE{{\cal E}}
\def\CF{{\cal F}}

\def\CH{{\cal H}}

\def\CM{{\cal M}}

\def\CO{{\cal O}}

\def\CT{{\cal T}}

\def\BH{\mathbb{H}}

\def\BZ{\mathbb{Z}}

\title{Fermionic and parafermionic CFTs with $\widehat{su}(2)$ and $\widehat{su}(3)$ symmetry}

\author{Kohki Kawabata}

\affiliation{Department of Physics, Faculty of Science,
The University of Tokyo,\\
Bunkyo-Ku, Tokyo 113-0033, Japan}

\abstract{
    We investigate two-dimensional conformal field theories (CFTs) with affine $\widehat{su}(2)$ and $\widehat{su}(3)$ algebra symmetry. Their bosonic modular-invariant partition functions have been fully classified based on the ADE classification. In this work, we extend the classification to include fermionic and parafermionic CFTs with the same affine symmetries, utilizing techniques of fermionization and parafermionization. We find that the fermionic and parafermionic $\widehat{su}(2)$ models are related to non-simply laced Dynkin diagrams.
}

\begin{document}
\maketitle
\flushbottom

\section{Introduction}

The classification of modular-invariant partition functions has been a central topic in two-dimensional conformal field theories (CFTs).
The spectrum of 2d bosonic theories is strongly constrained by the modular invariance of the torus partition function.
A classic result is the classification of modular-invariant partition functions for the Wess-Zumino-Witten (WZW) models with affine Lie algebra symmetries, such as $\widehat{su}(2)$ models~\cite{Cappelli:1986hf,Cappelli:1987xt,Kato:1987td,Gannon:1999cp}, $\widehat{su}(3)$ models~\cite{Gannon:1992ty,Gannon:1994cf}, level-one WZW models for simple  algebras~\cite{Degiovanni:1989ne,Gannon:1992nq}, and heterotic models~\cite{Gannon:1992np}.
In particular, the $\widehat{su}(2)$ models are fully classified based on simply laced Dynkin diagrams of the ADE type.
The correspondence between modular invariants and graphs has been extensively studied including the generalization to $\widehat{su}(3)$ models~\cite{DiFrancesco:1991st,Petkova:1995fw,Petkova:1996yv,Zuber:2000ia}.

A natural question is whether the classification of bosonic CFTs can be extended to fermionic and parafermionic CFTs.
When the theory is fermionic, it contains operators with half-integral spin and depends on the choice of spin structure of our spacetime. Correspondingly, the torus partition function is not invariant but covariantly transforms under modular transformation.\footnote{Toward the classification of fermionic rational CFTs, a modular bootstrap method has been recently developed in~\cite{Bae:2020xzl,Bae:2021mej,Duan:2022kxr}, which is closely related to the modular tensor categories (see e.g.~\cite{bruillard2017fermionic,Cho:2022kzf}).}
Among numerous constructions of fermionic CFTs, a recently proposed procedure called fermionization is a powerful tool for classification~\cite{Karch:2019lnn,Tachikawalec}.
Through fermionization, one can construct a fermionic CFT from a bosonic CFT with a global $\BZ_2$ symmetry.
Exploiting this technique has classified fermionic minimal models~\cite{Runkel:2020zgg,Hsieh:2020uwb, Kulp:2020iet} and chiral fermionic CFTs with relatively small central charges~\cite{BoyleSmith:2023xkd,Rayhaun:2023pgc,Hohn:2023auw}.
Similarly, for parafermionic theories, the torus partition functions depend on the choice of paraspin structure and intricately transform under modular transformation.
As well as fermionization, one can map a bosonic CFT with a global $\BZ_N$ symmetry to a parafermionic CFT via the so-called parafermionization.
Through the technique, the parafermionic analogs of minimal models have also been classified~\cite{Yao:2020dqx}.

This paper aims to classify fermionic and parafermionic CFTs with affine $\widehat{su}(2)$ and $\widehat{su}(3)$ symmetries at arbitrary levels.
Our approach utilizes the techniques of fermionization and parafermionization, which establish a one-to-one correspondence between a (para)fermionic theory and a bosonic theory with $\mathbb{Z}_N$ symmetry. 
Fermionization and parafermionization require us to ensure the absence of 't~Hooft anomaly for the $\BZ_N$ symmetry.
Notably, the global symmetries without anomaly have been enumerated for $\widehat{su}(2)$ and $\widehat{su}(3)$ modular invariants, under the assumption that the symmetry group preserves the affine algebra~\cite{Lienart:2000jw}.
Based on each identified non-anomalous symmetry, we construct corresponding fermionic and parafermionic CFTs, achieving a complete classification.

For $\widehat{su}(2)$ symmetry, each modular invariant labeled by ADE has a symmetry group equal to the automorphism group of the related Dynkin diagram~\cite{Lienart:2000jw}.
While many have only $\BZ_2$ symmetry, the theory related to the Dynkin diagram $D_4$ has the permutation group $S_3$, which contains a $\BZ_3$ subgroup.
We utilize the $\BZ_N$ group symmetries $(N=2,3)$ to construct fermionic and parafermionic theories with $\widehat{su}(2)$ symmetry.
We summarize the classification result of fermionic and parafermionic $\widehat{su}(2)$ theories in table~\ref{tab:listsu(2)}.
For $\widehat{su}(3)$ symmetry, most modular invariants have only $\BZ_3$ symmetry, but a theory called $\CD_6$ has an enhanced symmetry group equal to the alternating group $A_4$.
Correspondingly, in addition to $\BZ_3$ parafermionic theories, we have a fermionic theory obtained via fermionization of the $\BZ_2$ subgroup of the alternating group $A_4$.
We show the classification of $\widehat{su}(3)$ theories in table~\ref{tab:listsu(3)}.

Our results reveal that each fermionic and parafermionic $\widehat{su}(2)$ theory is related to a non-simply laced Dynkin diagram.
As in the ADE classification of $\widehat{su}(2)$ modular invariants, the relation is based on two facts:
\begin{enumerate}
    \item The height $n=k+2$ $(k:\mathrm{level})$ equals the Coxeter number.
    \item The diagonal terms in the partition function $Z[0,0]$ consist of the exponents.
\end{enumerate}
Here, $Z[0,0]$ denotes the torus partition function with spin structure $(\mathrm{NS},\mathrm{NS})$ for fermionic theories and its analog for parafermionic theories.
While fermionic CFTs labeled by $B_{\frac{n}{2}}$ and $C_{\frac{n}{2}}$ form a regular class consisting of an infinite number of theories, an exceptional class consists of the fermionic theory $F_4$ at height $n=12$ and the parafermionic theories $G_2$, $\overline{G_2}$ at height $n=6$.
Their relationships to bosonic $\widehat{su}(2)$ theories are shown in Fig.~\ref{fig:su2_ferm} and Fig.~\ref{fig:paraf_d4}, which suggests that the (para)fermionization map is related to the folding of Dynkin diagrams, a technique for constructing a non-simply laced diagram from a simply laced one by using its automorphism.
These observations show that the conventional relationship between bosonic $\widehat{su}(2)$ theories and simply-laced Dynkin diagrams is extended to non-simply laced cases when fermionic and parafermionic theories are included.

The organization of this paper is as follows.
Section~\ref{sec:modularinv} provides our convention and reviews modular invariants in WZW models with $\widehat{su}(2)$ and $\widehat{su}(3)$ symmetries.
In section~\ref{sec:jordanwigner}, we review the gauging of $\BZ_N$ global symmetry including orbifold, fermionization, and parafermionization.
We emphasize the relationship between an original bosonic theory and a fermionized or parafermionized theory.
The main part of this paper is section~\ref{sec:su(2)} and section~\ref{sec:su(3)}.
Through the gauging technique, we classify fermionic and parafermionic CFTs with $\widehat{su}(2)$ symmetries in section~\ref{sec:su(2)} and with $\widehat{su}(3)$ symmetries in section~\ref{sec:su(3)}.
We explicitly perform fermionization and parafermionization starting with the bosonic torus partition function graded by an element of a global symmetry.
We conclude in section~\ref{sec:discussion} with some discussions.

\section{Modular invariants in WZW models}
\label{sec:modularinv}

In this section, we review the modular-invariant torus partition functions of WZW models with $\widehat{su}(2)$ and $\widehat{su}(3)$ symmetries.
We emphasize that the theories with diagonal or non-diagonal modular invariants are both called WZW models.

Let $\CB$ be a two-dimensional bosonic CFT with an affine Lie algebra symmetry.
We take our spacetime as a torus with modulus $\tau = \tau_1+\i\tau_2$ valued in the upper half-plane $\BH$.
A torus has two independent cycles:
\begin{equation}
    \text{spacial:}\quad w\sim w+1\,,\qquad \text{temporal:}\quad w\sim w+\tau\,,
\end{equation}
where $w$ is a cylindrical coordinate on a torus.
Then, the torus partition function of the bosonic theory $\CB$ is given by
\begin{align}
    Z_\CB =  \sum_{p,\,p'} \,\chi_p \, \CM_{p,\,p'} \,\bar{\chi}_{p'}\,,
\end{align}
where $\chi_p:= \chi_p(\tau)$ is a character of a representation labeled by $p$.
Here, a mass matrix $\CM_{p,\,p'}$ consists of non-negative integers and glues holomorphic and anti-holomorphic characters.
Note that the partition function $Z_\CB$ depends on both $\tau$ and $\bar{\tau}$, although we omit them.

The torus partition function is invariant under modular transformations.
The general modular transformation forms a $\mathrm{PSL}(2,\BZ)$ group and can be expressed by
\begin{align}
\label{eq:modular}
    \tau \to \frac{a\tau+b}{c\tau+d}\,,\qquad a,b,c,d\in\BZ \quad\mathrm{and}\quad ad-bc=1\,.
\end{align}
In terms of an $\mathrm{SL}(2,\BZ)$ group, the transformation is denoted by a matrix $\left(\begin{smallmatrix}
    a & b \\ c & d
\end{smallmatrix}\right)\in \mathrm{SL}(2,\BZ)$.
Since a torus does not change under modular transformations, the torus partition function $Z_\CB$ should be invariant under modular transformations.

For a theory to be physical, we additionally require that each primary is included in the spectrum by only non-negative integers and the vacuum is unique.
One important task is to classify the physical partition functions satisfying modular invariance, the non-negativity of $\CM_{p,p'}$, and the uniqueness of the vacuum.
This work was completed for $\widehat{su}(2)$ WZW models in~\cite{Cappelli:1986hf,Cappelli:1987xt,Kato:1987td} and for $\widehat{su}(3)$ WZW models in~\cite{Gannon:1992ty,Gannon:1994cf}.
(See e.g., \cite{Mukhi:2022bte,Gowdigere:2023xnm} and references therein for recent developments in the classification of 2d rational CFTs with few primaries.)

\paragraph{Modular invariants of \texorpdfstring{$\widehat{su}(2)$}{su(2)} WZW models.}

Let us consider the affine $\widehat{su}(2)$ algebra with level $k$, i.e., the affine $\widehat{su}(2)_k$ algebra.
Below, we heavily use the height $n=k+2$ instead of the level $k$ for notational convenience.
Based on the Sugawara construction~\cite{Sugawara:1967rw}, the central charge is given by $c=3(n-2)/n$ and there are $n-1$ highest weight states $\ket{p}$ where $p$ is the shifted weight valued in $ P_{++}^{(n)}=\{1,2,\cdots,n-1\}$.
The trivial representation corresponds to $p=1$.
The conformal dimension of the highest weight state $\ket{p}$ is given by
\begin{align}
    h_p = \frac{p^2-1}{4n}\,.
\end{align}
Denoting the highest weight representation by $V_p$, we define its character by
\begin{align}
    \chi_p(\tau) = \Tr_{V_p} \left[\,q^{L_0-\frac{c}{24}}\,\right]\,,
\end{align}
where $L_0$ is the zero-mode of Virasoro generators and $q=e^{2\pi\i\tau}$.
Under the modular transformation, these characters transform as
\begin{equation}
    S:\;\chi_p(-1/\tau) = \sum_{p'} S_{p,p'}\,\chi_{p'}(\tau)\,,\qquad T:\; \chi_p(\tau+1) = \sum_{p'} T_{p,p'}\,\chi_{p'}(\tau)\,,
\end{equation}
where $S_{p,p'}$ and $T_{p,p'}$ are the modular transformation matrices
\begin{align}
    S_{p,p'} = \sqrt{\frac{2}{n}}\,\sin\left(\frac{\pi pp'}{n}\right)\,,\qquad
    T_{p,p'} = \delta_{p,p'}\,e^{2\pi\i (\frac{p^2}{4n}-\frac{1}{8})}\,,
\end{align}
for $p,p'\in P_{++}^{(n)}$.
Note that the charge conjugation $C=S^2$ is trivial for affine $\widehat{su}(2)$ algebras.

The modular-invariant partition functions of $\widehat{su}(2)$ WZW models are completely classified~\cite{Cappelli:1986hf,Cappelli:1987xt,Kato:1987td,Gannon:1999cp}.
The list of modular invariants is as follows: ($n=k+2)$
\begin{align*}
        A_{n-1}\; (n\geq3): \quad &\sum_{p=1}^{n-1}\;|\chi_p|^2\\
        
        D_{\frac{n}{2}+1} \; (n=4m+2):\quad &\sum_{p=1,\,\mathrm{odd}}^{2m-1}|\chi_p + \chi_{n-p}|^2+ 2\,|\chi_{2m+1}|^2\\
        
        D_{\frac{n}{2}+1}\; (n=4m+4):\quad &\sum_{p=1,\,\mathrm{odd}}^{n-1}|\chi_p|^2  + \sum_{p=2,\,\mathrm{even}}^{n-2} \chi_{p}\,\bar{\chi}_{n-p}\\
        
        E_6\;(n=12): \quad &|\chi_1+\chi_7|^2 + |\chi_4+\chi_8|^2 +|\chi_5+\chi_{11}|^2\\
        
        E_7\;(n=18): \quad &|\chi_1+\chi_{17}|^2 + |\chi_5+\chi_{13}|^2 +|\chi_7+\chi_{11}|^2 + |\chi_9|^2 
        + (\chi_9\,(\bar{\chi}_3+\bar{\chi}_{15})+\text{c.c.}) \\
        
        E_8\;(n=30): \quad &|\chi_1+\chi_{11} + \chi_{19} + \chi_{29}|^2 + |\chi_7 + \chi_{13} + \chi_{17} + \chi_{23}|^2
\end{align*}
The notable observation is that each modular invariant is related to a simply laced Dynkin diagram. The relation is based on the two facts:
\begin{enumerate}
    \item The height $n$ is equal to the Coxeter number of the corresponding Lie algebra.
    \item The set of $p\in P_{++}^{(n)}$ appearing in the diagonal terms $|\chi_p|^2$ of a modular invariant is exactly the set of exponents of the associated Dynkin diagram.
\end{enumerate}
We show the Coxeter number and exponents for each Dynkin diagram in table~\ref{tab:coxeter_extended}. Compare it with the above modular invariants.
We emphasize that the corresponding Lie algebra does not reflect a global symmetry of the modular invariant.

\begin{table}[t]
    \centering
    \renewcommand{\arraystretch}{1.2}
    \begin{tabular}{c c c} 
        \toprule
        \textbf{Type} & \textbf{Coxeter Number } & \textbf{Exponents} \\
        \midrule
        \( A_r \) & \( r+1 \) & \( 1, 2, \dots, r \) \\
        \( B_r \) & \( 2r \) & \( 1, 3, 5, \dots, 2r-1 \) \\
        \( C_r \) & \( 2r \) & \( 1, 3, 5, \dots, 2r-1 \) \\
        \( D_r \) & \( 2r-2 \) & \( 1, 3, 5, \dots, 2r-3, r-1 \) \\
        \( E_6 \) & \( 12 \) & \( 1, 4, 5, 7, 8, 11 \) \\
        \( E_7 \) & \( 18 \) & \( 1, 5, 7, 9, 11, 13, 17 \) \\
        \( E_8 \) & \( 30 \) & \( 1, 7, 11, 13, 17, 19, 23, 29 \) \\
        \( F_4 \) & \( 12 \) & \( 1, 5, 7, 11 \) \\
        \( G_2 \) & \( 6 \) & \( 1, 5 \) \\
        \bottomrule
    \end{tabular}
    \caption{Coxeter number and exponents for each Dynkin diagram, including both simply laced and non-simply laced types.}
    \label{tab:coxeter_extended}
\end{table}

\paragraph{Modular invariants of \texorpdfstring{$\widehat{su}(3)$}{su(3)} WZW models.}

Next, we consider the affine $\widehat{su}(3)$ algebra with level $k$, i.e., the affine $\widehat{su}(3)_k$ algebra.
We use the height $n=k+3$ for notational convenience.
Based on the Sugawara construction, the central charge is $c=8(n-3)/n$ and the highest weight states $\ket{p}$ are labeled by the shifted weights valued in 
\begin{equation}
P_{++}^{(n)} = \{p=(a,b)\in\BZ^2\mid 1\leq a,b,a+b\leq n-1\}\,.
\end{equation}
In this notation, the trivial representation is labeled by $p=(1,1)$.
The conformal dimension of the highest weight state $\ket{p=(a,b)}$ is given by
\begin{equation}
    h_{p} = \frac{a^2+ab+b^2-3}{3n}\,.
\end{equation}
As in the $\widehat{su}(2)$ case, we denote the character of the highest weight representation $V_p$ by $\chi_p(\tau)$.
The modular transformation matrices are given by
\begin{align}
\begin{aligned}
    S_{p,p'} &= \frac{-\i}{\sqrt{3}n}\,\sum_{w\,\in\, W(su(3))}\,(\mathrm{det}\, w) \,e^{-2\pi\i\frac{w(p)\cdot p'}{n}}\,,\\
    T_{p,p'} &= \delta_{p,p'}\,e^{2\pi\i \left(\frac{a^2+ab+b^2}{3n}-\frac{1}{3}\right)}\,,
\end{aligned}
\end{align}
where $p=(a,b)$ and $p'=(a',b')$, and $W(su(3))$ is the Weyl group of ${su}(3)$ algebra.
Note that the charge conjugation $C=S^2$ acts on a weight $p=(a,b)$ as $C(a,b)=(b,a)$.

The modular invariants of $\widehat{su}(3)$ models were considered in~\cite{Christe:1988vc,Moore:1988ss} and their completion was demonstrated in~\cite{Gannon:1992ty,Gannon:1994cf}.
The list of $\widehat{su}(3)$ modular invariants is the following: ($n=k+3$)
\begin{align*}
    \begin{aligned}
        \CA_{n}\; (n\geq4): \quad &\sum_{p}\;|\chi_p|^2\\
        
        \CD_{n}\; (n\notin 3\BZ):\quad &\sum_p \,  \chi_{\mu^{nt(p)}(p)}\,\bar{\chi}_p \\
        
        \CD_{n}\; (n\in 3\BZ):\quad &\sum_{t(p)=0\,\mathrm{mod}\,3}\left[\,|\chi_p|^2 + \chi_{\mu(p)}\,\bar{\chi}_p + \chi_{\mu^2(p)}\,\bar{\chi}_p\,\right]\\
        
        \CE_8\;(n=8): \quad &|\chi_{(1,1)}+\chi_{(3,3)}|^2+|\chi_{(3,1)}+\chi_{(3,4)}|^2 + |\chi_{(1,3)}+\chi_{(4,3)}|^2\\
        &\quad +|\chi_{(4,1)}+\chi_{(1,4)}|^2 + |\chi_{(2,3)}+\chi_{(6,1)}|^2+|\chi_{(3,2)}+\chi_{(1,6)}|^2\\
        
        \CE_{12}\;(n=12): \quad & |\chi_{(1,1)}+\chi_{(1,10)}+\chi_{(10,1)}+\chi_{(2,5)}+\chi_{(5,2)}+\chi_{(5,5)}|^2
        + 2\, |\chi_{(3,3)} +\chi_{(3,6)} + \chi_{(6,3)}|^2\\

        \CE_{12}'\;(n=12): \quad &
        |\chi_{(1,1)}+\chi_{(10,1)}+\chi_{(1,10)}|^2 + |\chi_{(3,3)}+\chi_{(3,6)} + \chi_{(6,3)}|^2 + 2\,|\chi_{(4,4)}|^2 \\
        &\quad
        +|\chi_{(1,4)}+\chi_{(7,1)} + \chi_{(4,7)}|^2 + |\chi_{(4,1)}+\chi_{(1,7)} + \chi_{(7,4)}|^2 \\
        &\quad
        + |\chi_{(5,5)}+\chi_{(5,2)} + \chi_{(2,5)}|^2  + (\chi_{(4,4)}\,(\bar{\chi}_{(2,2)}+\bar{\chi}_{(2,8)}+\bar{\chi}_{(8,2)})+\text{c.c.})\\
        
        \CE_{24}\;(n=24): \quad &
        |\chi_{(1,1)}+\chi_{(5,5)}+\chi_{(7,7)}+\chi_{(11,11)}+\chi_{(22,1)}+\chi_{(1,22)}   \\ &\quad + \chi_{(14,5)}+\chi_{(5,14)}+\chi_{(11,2)}+\chi_{(2,11)} + \chi_{(10,7)}+\chi_{(7,10)}|^2 \\ &\quad + 
        |\chi_{(16,7)}+\chi_{(7,16)}+\chi_{(16,1)}+\chi_{(1,16)}+\chi_{(11,8)}+\chi_{(8,11)} \\ &\quad +  \chi_{(11,5)}+\chi_{(5,11)}+\chi_{(8,5)}+\chi_{(5,8)} + \chi_{(7,1)}+\chi_{(1,7)}|^2
    \end{aligned}
\end{align*}
Here, $\mu$ is the order-three automorphism of $P_{++}^{(n)}$: $\mu(a,b) = (n-a-b,a)$, $\mu^2(a,b)=(b,n-a-b)$, $\mu^3(a,b)=(a,b)$.
Also, $t(a,b)=a+2b$ mod $3$ is called the triality of $p=(a,b)\in P_{++}^{(n)}$. Under the action of automorphism $\mu$, the triality $t$ transforms as $a+2b\to a+2b+n\to a+2b+2n$ mod $3$.
When $n\in3\BZ$, the triality is invariant under automorphism $\mu$.

In analogy with the $\widehat{su}(2)$ case, the $\widehat{su}(3)$ modular invariants are labeled by ADE.
Note that each modular invariant is not related to a Dynkin diagram, but to a generalized ADE diagram, which is constructed to extend the relationship between $\widehat{su}(2)$ modular invariants and graphs (see e.g. section 17.10 of~\cite{DiFrancesco:1997nk}).

Finally, we mention that the above modular invariants have a charge-conjugated partner by replacing the mass matrix $\CM_{p,p'}$ by $\CM_{p,C(p')}$, and some of them are invariant under the charge conjugation: $\CD_6 = \CD_6^\ast$, $\CD_9=\CD_9^\ast$, $\CE_{12}=\CE_{12}^\ast$, and $\CE_{24} =\CE_{24}^\ast$.

\section{Orbifold, fermionization, and parafermionization}
\label{sec:jordanwigner}

This section briefly reviews the orbifold, fermionization, and parafermionization of a 2d bosonic CFT with $\BZ_N$ global symmetry.
These transformations are a generalized version of Jordan-Wigner transformation~\cite{Schultz:1964fv,Fradkin:1980th} and can be described as gauging a bosonic theory with a non-anomalous $\BZ_N$ symmetry.
Throughout this paper, we collectively refer to fermionization and parafermionization as the generalized Jordan-Wigner transformation.

Let $\CB$ be a two-dimensional bosonic theory with a non-anomalous $\BZ_N$ symmetry $G=\{1,g,g^2,\cdots,g^{N-1}\}$. 
One can insert the background $\BZ_N$ fields on our spacetime and compute the corresponding partition functions.
The torus partition functions twisted by the elements $(g^{a_1},g^{a_2})$ are given by
\begin{align}
    Z_\CB[a_1,a_2] = \Tr_{\CH_{a_1}}\left[\,g^{a_2} q^{L_0 - \frac{c}{24}}\,\bar q^{\bar L_0 - \frac{c}{24}}\,\right] \ ,
\end{align}
where $\CH_{a}$ denotes the Hilbert space quantized under the $g^a$-twisted periodicity: $\phi(x+2\pi)=g^a\cdot \phi(x)$ for a bosonic field $\phi$ with a unit charge.
The twisted partition functions transform under the modular transformation by (see e.g.~\cite{Polchinski:1998rq})
\begin{align}
\label{eq:twisted_modular}
    S:\; Z_\CB[a_1,a_2] \to Z_\CB[a_2,-a_1]\,,\qquad T:\; Z_\CB[a_1,a_2] \to Z_\CB[a_1,a_1+a_2]\,,
\end{align}
where we used the definition of the modular $S$ transformation $\bigl(\begin{smallmatrix}
    0 & -1 \\ 1 & 0
\end{smallmatrix}\bigr)$ and modular $T$ transformation $\bigl(\begin{smallmatrix}
    1 & 1 \\ 0 & 1
\end{smallmatrix}\bigr)$ in the general expression~\eqref{eq:modular}.

\paragraph{Orbifold.}
Orbifolding by the non-anomalous $\BZ_N$ symmetry constitutes a new consistent bosonic theory $\CO=\CB/G$~\cite{Dixon:1985jw,Dixon:1986jc,Dixon:1986qv}. 
The orbifold theory does not have the original $\BZ_N$ symmetry $G$ because it is gauged and becomes trivial in the new theory $\CO$.
Instead, a dual $\BZ_N$ symmetry $\widehat{G}$ emerges after orbifolding~\cite{Vafa:1989ih,Bhardwaj:2017xup}.
We can insert the background fields associated with $\widehat{G}$ on a torus.
The twisted partition functions of the orbifold theory $\CO$ are given by
\begin{align}
\label{eq:gene_orb}
    Z_\CO[\hat{a}_1,\hat{a}_2] = \frac{1}{N}\sum_{a} \,\omega_N^{-(\hat{a}_1 a_2-\hat{a}_2a_1)}\, Z_\CB[a_1,a_2]\,,
\end{align}
where $\omega_N = \exp(2\pi\i/N)$.
The untwisted partition function $Z_\CO[0,0]$ is still invariant under modular transformations.
This indicates that the orbifold theory $\CO$ has a consistent spectrum by itself.
Note that if the original symmetry $G$ has an 't~Hooft anomaly, then the orbifold theory is not invariant under modular transformations, which implies we can diagnose the anomalies by modular transformations on a torus~\cite{Numasawa:2017crf,Lin:2019kpn,Lin:2021udi}.
We can further take the orbifold by the dual symmetry $\widehat{G}$, which results in the original theory $\CO/\widehat{G}=\CB$.

\paragraph{Fermionization.}

A fermionic theory includes a half-integral spin in its spectrum and requires a choice of the spin structure of spacetime.
When the spacetime is a torus, one can specify its spin structure by the periodicity of fermionic fields $\psi$ along the two independent cycles. A choice of spin structure can be denoted by $s=(s_1,s_2)$ where $s_i\in \{0,1\}$.
For $s_i=0$, fermionic fields $\psi$ are anti-periodic, while, for $s_i=1$, they are periodic.
The Hilbert space quantized under the anti-periodic boundary condition $\psi\to -\psi$ is called the NS sector $\CH_\mathrm{NS}$, and the one under the periodic boundary condition $\psi\to \psi$ is called the R sector $\CH_\mathrm{R}$.

The partition functions of a fermionic theory $\CF$ depend on the choice of spin structure.
For a spin structure $(s_1,s_2)$, the torus partition functions are given by
\begin{align}
    Z_\CF[s_1,s_2] = \Tr_{\CH_{\iota(s_1)}}\left[\,(-1)^{s_2F} q^{L_0-\frac{c}{24}}\,\bar{q}^{\bar{L}_0-\frac{c}{24}}\,\right]\,,
\end{align}
where $\iota: \{0,1\}\to \{\mathrm{NS},\mathrm{R}\}$ such that $\iota(0) = \mathrm{NS}$ and $\iota(1)=\mathrm{R}$.
We call $Z_\CF[0,0]$ the NS-NS partition function, $Z_\CF[0,1]$ the NS-R partition function, and so on.

Unlike bosonic partition functions, the fermionic partition functions covariantly transform under the modular transformation.
The transformation rules are given by (\cite{Ginsparg:1988ui})
\begin{align}
    T: \;Z_\CF[s_1,s_2]\to Z_\CF[s_1,1+s_1+s_2]\,,\qquad S:\; Z_\CF[s_1,s_2] \to Z_\CF[-s_2,s_1]\,.
\end{align}
The NS-NS partition function is invariant under the modular transformations generated by $S$ and $T^2$, while the R-R partition function is invariant under the full $\mathrm{SL}(2,\BZ)$ group.

A bosonic theory with a non-anomalous $\BZ_2$ symmetry can be mapped to a fermionic one by fermionization~\cite{Karch:2019lnn,Tachikawalec}.
See~\cite{Runkel:2020zgg,Fukusumi:2023psx} for an algebraic description and~\cite{Ji:2019ugf,Bae:2021lvk,Kikuchi:2022jbl,Fukusumi:2022ucr} for its applications.
This utilizes the low-energy limit of a non-trivial fermionic topological phase described by the Kitaev-Majorana chain~\cite{Kitaev:2000nmw}.
We call the spin topological theory the Arf theory.
The partition function of the Arf theory under the spin structure $s=(s_1,s_2)$ is
\begin{align}
    Z_{\mathrm{Kitaev}}[s_1,s_2] = (-1)^{\mathrm{Arf}[s]}\,,
\end{align}
where $\mathrm{Arf}[s]$ shows the Arf invariant.
On a torus with spin structure $s=(s_1,s_2)$, the Arf invariant becomes
\begin{align}
    \mathrm{Arf}[s] 
        = s_1 s_2\,.
\end{align}
Namely, the Arf invariant is $-1$ for $s=(1,1)$ and $+1$ otherwise.
By coupling the Kitaev-Majorana chain to the original bosonic theory $\CB$ through a non-anomalous $\BZ_2$ symmetry, the bosonic theory $\CB$ turns into the fermionic one $\CF$:
\begin{align}
    \CF = \frac{\CB\times (-1)^{\mathrm{Arf}}}{\BZ_2}\,.
\end{align}
Correspondingly, the fermionized partition functions are 
\begin{align}
\label{eq:ferm_map}
    Z_\CF[s_1,s_2] = \frac{1}{2}\sum_{a}\, (-1)^{\mathrm{Arf}[s+a]} \,Z_\CB[a_1,a_2]\,. 
\end{align}
The inverse transformation gives the bosonization map (GSO projection)
\begin{align}
    Z_\CB[a_1,a_2] = \frac{1}{2} \sum_s \, (-1)^{\mathrm{Arf}[s+a]} \,Z_\CF[s_1,s_2]\,.
\end{align}

We can start with the orbifold theory $\CO$ whose partition functions are \eqref{eq:gene_orb} with $N=2$.
When the dual $\BZ_2$ symmetry $\widehat{G}$ is non-anomalous, this bosonic theory $\CO$ can be fermionized into another theory $\widetilde{F}$.
One can show that the two fermionized theories are related by stacking the Arf theory:
\begin{align}
    Z_{\widetilde{\CF}}[s_1,s_2] = (-1)^{\mathrm{Arf}[s]}\,Z_\CF[s_1,s_2]\,.
\end{align}
We summarize the relationships between orbifold and fermionization in Fig.~\ref{fig:commutative_ferm}.
The stacking of the Arf theory is converted into $\BZ_2$-orbifolding via bosonization/fermionization map.
When a bosonic theory is self-dual under orbifolding, the associated fermionic theory is invariant under stacking the Arf theory.

\begin{figure}[tbp]
    \centering
    \begin{tikzpicture}[transform shape]
    \draw (0,0)node[]{$\CO$};
    \draw[<->,>=stealth] (3.75,-0.0) to node[below,font=\small]{bosonize} node[above,font=\small]{fermionize}(0.75,-0.0);
    \draw (4.5,0)node[]{$\widetilde{\CF}$};
    \draw (0,3)node[]{$\CB$};
    \draw[<->,>=stealth] (0.75,3) to node[below,font=\small]{bosonize} node[above,font=\small]{fermionize} (3.75,3.0);
    \draw (4.5,3) node[]{$\CF$};
    \draw[<->,>=stealth] (4.5,0.5) to node[right,font=\small]{stack} (4.5,2.5);
    \draw[<->,>=stealth] (0,0.5) to node[left,font=\small]{orbifold} (0,2.5);
    \end{tikzpicture}
    \caption{Fermionization/bosonization map associated with a non-anomalous $\BZ_2$ symmetry.}
    \label{fig:commutative_ferm}
\end{figure}

\paragraph{Parafermionization.}

A parafermionic theory has a fractional spin in its spectrum~\cite{Fradkin:1980th,Fateev:1987vh,Gepner:1986hr} and requires choosing a paraspin structure of our spacetime.
Since a paraspin structure is not fully understood beyond a torus, we set our spacetime as a torus.
For a $\BZ_N$ parafermionic theory, a paraspin structure on a torus is specified by the periodicity $(k_1,k_2)\in\BZ_N\times\BZ_N$ of parafermionic fields along the two independent cycles.

We can construct a parafermionic theory from a bosonic theory with a non-anomalous $\BZ_N$ symmetry via parafermionization~\cite{Yao:2020dqx} (see also~\cite{Thorngren:2021yso,Burbano:2021loy,Fukusumi:2022kly,Duan:2023ykn}).
As in fermionization, parafermionization can be described by gauging a bosonic theory with the low-energy limit of a parafermionic chain~\cite{alexandradinata2016parafermionic}.
The torus partition functions are given by
\begin{align}
    Z_{\mathrm{TQFT}}[k_1,k_2] = \omega_N^{\mathrm{Arf}_N[k]}\,,
\end{align}
where $\mathrm{Arf}_N[k]$ represents one generalization of the $\BZ_2$-valued Arf invariant to the $\BZ_N$ case. On a torus with paraspin structure $k=(k_1,k_2)$, 
\begin{align}
    \mathrm{Arf}_N[k] = k_1k_2\mod N\,.
\end{align}
By coupling the parafermionic chain to a bosonic theory $\CB$ and gauging the diagonal $\BZ_N$ symmetry, we obtain a parafermionic theory
\begin{align}
    \mathrm{PF} = \frac{\CB\times \omega_N^{\mathrm{Arf}_N}}{\BZ_N}\,.
\end{align}
The $\BZ_N$ parafermionized partition functions are given by
\begin{align}
    Z_{\mathrm{PF}_\rho}[k_1,k_2] = \frac{1}{N} \sum_{a_1,\,a_2\,\in\,\BZ_N} \omega_N^{\rho\, (k_1+a_1)(k_2+a_2)} Z_\CB[a_1,a_2]\,,
\end{align}
where a parameter $1\leq\rho< N$ is coprime with $N$.
In general, $\rho$ is valued in the automorphism group of $\BZ_N$, a multiplicative group $\BZ_N^\times$, whose order is Euler's totient function $\phi(N)$.
When $N=2$, the expression reproduces the fermionization map~\eqref{eq:ferm_map}.
Under the modular transformation $\tau\to (a\tau+b)/(c\tau+d)$, the parafermionic partition functions transform as
\begin{align}
    Z_{\mathrm{PF}_\rho}[k_1,k_2] \to  \sum_{k_1',k_2'}\, M_{k_1,k_2}^{k_1',k_2'} \,Z_{\mathrm{PF}_\rho}[k_1',k_2']\,,
\end{align}
where a matrix $M$ is a representation of $\left(\begin{smallmatrix}
    a & b\\ c&d
\end{smallmatrix}\right)\in \mathrm{SL}(2,\BZ)$ given  by
\begin{align}
    M_{k_1,k_2}^{k_1',k_2'} = \frac{1}{N^2}\sum_{\ell_1,\ell_2} \,(\omega_N^\rho)^{\mathrm{Arf}_N[k+\ell']-\mathrm{Arf}_N[k'+\ell]}\,.
\end{align}
Here, $\ell_i,\ell_i'\in\BZ_N$ and $\ell' = (\ell_1',\ell_2') = (a\ell_1-c\ell_2,-b\ell_1+d\ell_2)$.
This shows that the $\BZ_N$-parafermionized partition functions do not transform covariantly when $N\geq 3$~\cite{Yao:2020dqx,Duan:2023ykn}.

The inverse of this procedure yields the bosonization map
\begin{align}
     Z_\CB[a_1,a_2] = \frac{1}{N} \sum_{k_1,\,k_2\,\in\,\BZ_N} \omega_N^{-\rho\, (k_1+a_1)(k_2+a_2)} Z_{\mathrm{PF}_\rho}[k_1,k_2]\,.
\end{align}
Also, starting with the orbifold theory $\CO$, one can obtain another parafermionic theory $\widetilde{\mathrm{PF}}$.
The two parafermionic theories are related by
\begin{align}
\label{eq:para_orb}
    Z_{\widetilde{\mathrm{PF}}_\rho}[k_1,k_2] = \omega_N^{\rho \,k_1k_2}\,Z_{\mathrm{PF}_{1/\rho}}[-\rho k_1,\rho k_2]\,.
\end{align}
Thus, the two parafermionic theories are exchanged by the combined operation of stacking the $\mathrm{Arf}_N$ theory and conjugation $(k_1,k_2)\to (-\rho k_1,\rho k_2)$ (see Fig.~\ref{fig:commutative_paraferm}).
Note that the two parafermionic theories are identical if the bosonic theory $\CB$ is self-dual under the $\BZ_N$ orbifold.

Finally, we comment on the $\BZ_3$ parafermionization, which is the only case we need in the later sections.
For $\BZ_3$ symmetry, the parameter $\rho$ takes values $\rho=\pm1$ mod $3$.
For $\rho=1$, the relation \eqref{eq:para_orb} simplifies to
\begin{align}
\label{eq:relpara_1}
    Z_{\widetilde{\mathrm{PF}}_1}[k_1,k_2] = \omega_N^{k_1k_2}\,Z_{\mathrm{PF}_1}[-k_1,k_2]\,.
\end{align}
For $\rho=-1=2$ mod $3$, this becomes 
\begin{align}
\label{eq:relpara_2}
    Z_{\widetilde{\mathrm{PF}}_2}[k_1,k_2] = \omega_N^{-k_1k_2}\,Z_{\mathrm{PF}_2}[k_1,-k_2]\,.
\end{align}
Note that the theories with and without tilde have the same partition function when the paraspin structure is $(k_1,k_2)=(0,0)$.

\begin{figure}[tbp]
    \centering
    \begin{tikzpicture}[transform shape]
    \draw (0,0)node[]{$\CO$};
    \draw[<->,>=stealth] (3.75,-0.0) to node[below,font=\small]{bosonize} node[above,font=\small]{parafermionize}(0.75,-0.0);
    \draw (4.5,0)node[]{$\widetilde{\mathrm{PF}}$};
    \draw (0,3)node[]{$\CB$};
    \draw[<->,>=stealth] (0.75,3) to node[below,font=\small]{bosonize} node[above,font=\small]{parafermionize} (3.75,3.0);
    \draw (4.5,3) node[]{${\mathrm{PF}}$};
    \draw[<->,>=stealth] (4.5,0.5) to node[right,font=\small,align=left]{stack\\+conjugate} (4.5,2.5);
    \draw[<->,>=stealth] (0,0.5) to node[left,font=\small]{orbifold} (0,2.5);
    \end{tikzpicture}
    \caption{Parafermionization/bosonization map associated with a non-anomalous $\BZ_N$ symmetry for a given $\rho\in \BZ_N^\times$.}
    \label{fig:commutative_paraferm}
\end{figure}

\section{Classification of \texorpdfstring{$\widehat{su}(2)$}{su(2)} models}
\label{sec:su(2)}

In this section, we classify the fermionic and parafermionic CFTs with affine $\widehat{su}(2)$ algebra symmetry.
As discussed in section~\ref{sec:modularinv}, the ADE classification completes the list of bosonic modular invariants~\cite{Cappelli:1986hf,Kato:1987td,Gannon:1992ty}.
Our strategy for classification is based on the generalized Jordan-Wigner transformation of the bosonic modular invariants by non-anomalous $\BZ_N$ symmetries, which was applied to the classification of fermionic and parafermionic minimal models ~\cite{Hsieh:2020uwb,Kulp:2020iet,Yao:2020dqx} and chiral fermionic CFTs~\cite{BoyleSmith:2023xkd,Rayhaun:2023pgc,Hohn:2023auw}.

The basic logic is as follows: Let us take a fermionic or parafermionic theory.
We can always bosonize the theory to obtain a bosonic theory $\CB$ with a non-anomalous $\BZ_N$ symmetry.
Namely, a fermionic and parafermionic CFT has a one-to-one correspondence with a bosonic theory $\CB$ with a non-anomalous $\BZ_N$ symmetry:
\begin{align}
\begin{aligned}
    \CB \text{ with $\BZ_2$ symmetry} \quad &\longleftrightarrow \quad \text{Fermionic theory }\CF\,,\\
    \CB \text{ with $\BZ_N$ symmetry} \quad &\longleftrightarrow \quad \text{Parafermionic theory }\mathrm{PF}\,.
\end{aligned}
\end{align}
On the other hand, the authors of~\cite{Lienart:2000jw} determine the non-anomalous finite symmetries, which preserves the affine algebra, in $\widehat{su}(2)$ and $\widehat{su}(3)$ modular invariants.
Thus, for each $\BZ_N$ symmetry of bosonic theories, we can perform the generalized Jordan-Wigner transformation and obtain the classification of fermionic and parafermionic theories.

Before proceeding to the detailed construction of each theory, we summarize the classification results of fermionic and parafermionic $\widehat{su}(2)$ models.
We show the list of fermionic and parafermionic theories with $\widehat{su}(2)_k$ symmetry at an arbitrary level $k$ in table~\ref{tab:listsu(2)}.
The list provides fermionic and parafermionic torus partition functions for each theory.
We can see that they are related to Dynkin diagrams based on two facts: (1) The height $n=k+2$ is equal to the Coxeter number. (2) The diagonal terms in the partition function with (para)spin structure $(0,0)$ consist of the exponents.
One can check these from table~\ref{tab:coxeter_extended}.
Except for $G_2$, we can see a one-to-one correspondence between $\widehat{su}(2)$ theories and non-simply Dynkin diagrams.
The relationship demonstrates a generalization of the ADE classification to non-simply laced BCFG diagrams.
Thus, bosonic, fermionic, and parafermionic theories with $\widehat{su}(2)$ symmetry are classified following both simply laced and non-simply laced Dynkin diagrams.

The $B$-type and $C$-type fermionic theories are related to the $A$-type and $D$-type bosonic theories through the left of Fig.~\ref{fig:su2_ferm}.
Note that the Dynkin diagrams $B_r$ and $C_r$ have the same Coxeter number and exponents at a fixed $r$. Thus, there is the freedom to swap their names. The only difference between $B$-type and $C$-type is the stacking of the Arf theory.
In particular, for $n=4$, the two theories $B_2$ and $C_2$ coincide. This is consistent with the accidental isomorphism of the Dynkin diagrams: $B_2\cong C_2$.
The fermionic theory $F_4$ is the fermionization of the modular invariant $E_6$ as in the right of Fig.~\ref{fig:su2_ferm}.
The parafermionic theories $G_2$ and $\overline{G_2}$ are the $\BZ_3$ parafermionization of the modular invariant $D_4$ with $\rho=1$ and $\rho=2$, respectively (see Fig.~\ref{fig:paraf_d4}).
We observe that these relationships are quite similar to the folding of Dynkin diagrams, which we will discuss in section~\ref{sec:discussion}.

\begin{table}[t]
    \centering
    \begin{tabular*}{\textwidth}{c@{\hskip 0.8cm}l}
    \toprule\addlinespace
         \textbf{Name} &  \textbf{Fermionic or parafermionic partition functions}\\
         \addlinespace\midrule\addlinespace

          \begin{tabular}{c}
          $B_{\frac{n}{2}}$ \\($n=4m+4$)
          \end{tabular}
          & \small{$\displaystyle Z_{\CF}[0,0]=\sum_{p=1,\,\mathrm{odd}}^{\frac{n}{2}-1}|\chi_p+\chi_{n-p}|^2$\, \quad $\displaystyle Z_{\CF}[1,1] = \sum_{p=2,\,\mathrm{even}}^{\frac{n}{2}-2}|\chi_p-\chi_{n-p}|^2$} \\ \addlinespace\addlinespace

          \begin{tabular}{c}
          $B_{\frac{n}{2}}$ \\($n=4m+2$)
          \end{tabular}
          & 
          \begin{tabular}{c}
          \small{$\displaystyle Z_{\CF}[0,0]=\sum_{p=1,\,\mathrm{odd}}^{n-1} |\chi_p|^2 + \sum_{p=2,\,\mathrm{even}}^{n-2} \chi_{n-p}\,\bar{\chi}_p$}\, \\
          
          \small{$\displaystyle Z_{\CF}[1,1] = \sum_{p=2,\,\mathrm{even}}^{n-2} |\chi_p|^2-\sum_{p=1,\,\mathrm{odd}}^{n-1} \chi_{n-p}\,\bar{\chi}_p $}
          \end{tabular}
          \\ \addlinespace\midrule[0.03em]\addlinespace

          \begin{tabular}{c}
            $C_{\frac{n}{2}}$ \\($n=4m+4$)
            \end{tabular}
            & \small{$\displaystyle Z_{\CF}[0,0]=\sum_{p=1,\,\mathrm{odd}}^{\frac{n}{2}-1}|\chi_p+\chi_{n-p}|^2$\, \quad $\displaystyle Z_{\CF}[1,1] = -\sum_{p=2,\,\mathrm{even}}^{\frac{n}{2}-2}|\chi_p-\chi_{n-p}|^2$} \\ \addlinespace\addlinespace
  
            \begin{tabular}{c}
            $C_{\frac{n}{2}}$ \\($n=4m+2$)
            \end{tabular}
            & 
            \begin{tabular}{c}
            \small{$\displaystyle Z_{\CF}[0,0]=\sum_{p=1,\,\mathrm{odd}}^{n-1} |\chi_p|^2 + \sum_{p=2,\,\mathrm{even}}^{n-2} \chi_{n-p}\,\bar{\chi}_p$}\, \\
            
            \small{$\displaystyle Z_{\CF}[1,1] =  \sum_{p=1,\,\mathrm{odd}}^{n-1} \chi_{n-p}\,\bar{\chi}_p  - \sum_{p=2,\,\mathrm{even}}^{n-2} |\chi_p|^2$}
            \end{tabular}
            \\ \addlinespace \midrule[0.03em]\addlinespace[0.4cm]

          \begin{tabular}{c}$F_{4}$ \;  ($n=12$)\end{tabular} & \small{$\displaystyle Z_{\CF}[0,0]=|\chi_1+\chi_5+\chi_7+\chi_{11}|^2$\,\qquad $Z_{\CF}[1,1]=0$}\\ \addlinespace[0.4cm]\midrule[0.07em]\addlinespace[0.4cm]

           $G_{2}$ \; ($n=6$) & \small{$\displaystyle Z_\mathrm{PF}[0,0]=|\chi_1+\chi_5|^2 +2\,(\chi_1+{\chi}_5) \,\bar{\chi}_3$}\\ \addlinespace[0.2cm]\addlinespace

           $\overline{G_{2}}$ \; ($n=6$) & \small{$\displaystyle Z_\mathrm{PF}[0,0]=|\chi_1+\chi_5|^2 +2\, \chi_3\,(\bar{\chi}_1+\bar{\chi}_5)$}\\ \addlinespace[0.4cm]\bottomrule
          
    \end{tabular*}
    \caption{List of fermionic and parafermionic CFTs with affine $\widehat{su}(2)$ algebra symmetry at each height $n=k+2$. We label each theory using the related Dynkin diagram. For the fermionic theories $B_{\frac{n}{2}}$, $C_{\frac{n}{2}}$, and $F_4$, we show the NS-NS partition function $Z_\CF[0,0]$ and R-R partition function $Z_\CF[1,1]$. For the $\BZ_3$ parafermionic theories $G_2$ and $\overline{G_2}$, we show the partition function $Z_{\mathrm{PF}}[0,0]$ with the paraspin structure $(k_1,k_2)=(0,0)$.}
    \label{tab:listsu(2)}
\end{table}

\subsection{Fermionic models from \texorpdfstring{$A_{n-1}$}{A{n-1}} and \texorpdfstring{$D_{\frac{n}{2}+1}$}{D{n/2+1}}}
Let us consider the fermionization of the diagonal modular invariants $A_{n-1}$, which has a $\BZ_2$ symmetry.
Since the $\BZ_2$ symmetry is anomalous unless the level $k$ is even, we focus on $n=k+2\in2\BZ$.
The twisted partition functions of the bosonic theory are given by
\begin{align}
\begin{aligned}
\label{eq:Z2su2a}
    Z_{A_{n-1}}[0,0] &=\sum_{p=1}^{n-1} \,|\chi_p|^2\,,\qquad\;\;\;\; Z_{A_{n-1}}[0,1]=\sum_{p=1}^{n-1} \,(-1)^{p+1}\,|\chi_p|^2\,,\\
    Z_{A_{n-1}}[1,0] &= \sum_{p=1}^{n-1} \,\chi_{n-p}\,\bar{\chi}_p\,,\qquad
    Z_{A_{n-1}}[1,1]= \sum_{p=1}^{n-1} \,(-1)^{p+n/2}\,\chi_{n-p}\,\bar{\chi}_p\,.
\end{aligned}
\end{align}
Due to the last partition function $Z_{A_{n-1}}[1,1]$, the action of the $\BZ_2$ symmetry depends on whether $n/2$ is even or odd, and the corresponding orbifold yields the modular invariant $D_{\ell}$ $(\ell:\mathrm{even})$ for $n\in 4\BZ+2$ and $D_\ell$ $(\ell:\mathrm{odd})$ for $n\in4\BZ$.

Let us consider the fermionization of the diagonal modular invariant by the $\BZ_2$ symmetry~\eqref{eq:Z2su2a}.
For $n=4m+4$, the fermionized theory of the modular invariant $A_{n-1}$ has the following partition functions:
\begin{align}
    \begin{aligned}
    \label{eq:btype-even}
        Z_{\CF, \,A_{n-1}}[0,0] &= \sum_{p=1,\,\mathrm{odd}}^{\frac{n}{2}-1}|\chi_p+\chi_{n-p}|^2\,, &
        Z_{\CF, \,A_{n-1}}[0,1] &= \sum_{p=1,\,\mathrm{odd}}^{\frac{n}{2}-1}|\chi_p-\chi_{n-p}|^2\,,\\
        Z_{\CF, \,A_{n-1}}[1,0] &= \sum_{p=2,\,\mathrm{even}}^{\frac{n}{2}-2}|\chi_p+\chi_{n-p}|^2 + 2\,|\chi_{\frac{n}{2}}|^2\,, &
        Z_{\CF, \,A_{n-1}}[1,1] &= -\sum_{p=2,\,\mathrm{even}}^{\frac{n}{2}-2}|\chi_p-\chi_{n-p}|^2\,.
    \end{aligned}
\end{align}
These fermionic partition functions are block-diagonal.
Note that the diagonal terms $|\chi_p|^2$ of the NS-NS partition function consist of $p=1,3,\cdots,4m+3$.
Notably, this set is the same as the set of the exponents of the Dynkin diagram $C_\frac{n}{2}$ where $n=4m+4$.
Additionally, the height $n$ is the Coxeter number of the Dynkin diagram $C_{\frac{n}{2}}$.
Following the ADE classification of modular invariants, we denote by $C_{\frac{n}{2}}$ the fermionic partition functions~\eqref{eq:btype-even}.

Also, we can start with the modular invariant $D_{\frac{n}{2}+1}$ ($n=4m+4$). Then, the resulting fermionic theory has the partition functions
\begin{align}
\label{eq:ctype}
    Z_{\CF,\,D_{\frac{n}{2}+1}}[s_1,s_2] = (-1)^{s_1s_2}\,Z_{\CF,\,A_{n-1}}[s_1,s_2]\,,
\end{align}
which are the same as \eqref{eq:btype-even} up to the sign of the R-R partition function. 
This sign flip can be interpreted as the stacking of the Arf theory.
Since the NS-NS partition function is the same as \eqref{eq:btype-even}, its diagonal terms are also $p=1,3,\cdots,4m+3$.
This time, we propose to call this fermionic theory $B_{\frac{n}{2}}$.
One reason is that the Dynkin diagram $B_{\frac{n}{2}}$ has the exponents $1,3,\cdots,4m+3$ and the Coxeter number $n$.
The other reason is that we can observe the accidental isomorphism $B_2\cong C_2$ in the associated fermionic theories:
\begin{align}
    Z_{C_{2}}[s_1,s_2] = Z_{B_2}[s_1,s_2]\,,
\end{align}
where we renamed $Z_{\CF,\,A_{3}}$ as $Z_{C_2}$ and $Z_{\CF,\,D_{3}}$ as $Z_{B_2}$.
This follows from that these theories are equivalent to 3 copies of Majorana fermions with the vanishing R-R partition function $Z_{B_2}[1,1] = Z_{C_2}[1,1]=0$.
Hence, the fermionized theories of the $D$-type modular invariants can be denoted by $B_{\frac{n}{2}}$.
Note that the difference between $B_{\frac{n}{2}}$ and $C_{\frac{n}{2}}$ theories is only stacking the Arf theory: $B_{\frac{n}{2}} = (-1)^{\mathrm{Arf}}\,C_{\frac{n}{2}}$.

For $n=4m+2$, we can similarly construct fermionic partition functions. The fermionized theory of modular invariant $A_{n-1}$ has the torus partition functions
\begin{align}
\begin{aligned}
\label{eq:btype-odd}
    Z_{\CF,\, A_{n-1}}[0,0] = \sum_{p=1,\,\mathrm{odd}}^{n-1} |\chi_p|^2 + \sum_{p=2,\,\mathrm{even}}^{n-2} \chi_{n-p}\,\bar{\chi}_p\,,\\
    Z_{\CF,\, A_{n-1}}[0,1] = \sum_{p=1,\,\mathrm{odd}}^{n-1} |\chi_p|^2 - \sum_{p=2,\,\mathrm{even}}^{n-2} \chi_{n-p}\,\bar{\chi}_p\,,\\
    Z_{\CF,\, A_{n-1}}[1,0] = \sum_{p=2,\,\mathrm{even}}^{n-2} |\chi_p|^2 +  \sum_{p=1,\,\mathrm{odd}}^{n-1} \chi_{n-p}\,\bar{\chi}_p\,,\\
    Z_{\CF,\, A_{n-1}}[1,1] =  \sum_{p=1,\,\mathrm{odd}}^{n-1} \chi_{n-p}\,\bar{\chi}_p- \sum_{p=2,\,\mathrm{even}}^{n-2} |\chi_p|^2\,.
\end{aligned}
\end{align}
In the NS-NS partition function $Z_{\CF,\, A_{n-1}}[0,0]$, the second term does not contribute to the diagonal terms $|\chi_p|^2$ since $n\in4\BZ+2$.
Thus, the diagonal terms of the NS-NS partition function consist of $p=1,3,\cdots,4m+1$.
As in $n=4m+4$, we propose to name the fermionic partition functions \eqref{eq:btype-odd} as $C_{\frac{n}{2}}$.
Also, the fermionization of modular invariant $D_{\frac{n}{2}+1}$ with $n=4m+2$ yields the partition functions related by \eqref{eq:ctype}.
Similarly, we call this theory $B_{\frac{n}{2}}$.

We summarize the relationship between bosonic modular invariants ($A$-type and $D$-type) and their fermionization ($B$-type and $C$-type) in the left of Fig.~\ref{fig:su2_ferm}.
Under the $\BZ_2$ orbifolding, the modular invariant $A_{n-1}$ is related to $D_{\frac{n}{2}+1}$.
After fermionization, this relation is encoded into the stacking of the Arf theory.

Finally, we mention that the fermionization should be related to the folding of the corresponding Dynkin diagram. When a simply laced Dynkin diagram has an automorphism, one can obtain a non-simply laced Dynkin diagram by an operation called folding (see Fig.~5.3 in~\cite{stekolshchik2005notes}). The Dynkin diagrams $A_{n-1}$ and $D_{\frac{n}{2}+1}$ can be folded only when $n$ is even and yield $C_{\frac{n}{2}}$ and $B_{\frac{n}{2}}$, respectively. This exactly agrees with our fermionization map in Fig.~\ref{fig:su2_ferm}.

\begin{figure}
    \centering
    \begin{minipage}{0.48\textwidth}
    \begin{tikzpicture}[transform shape, scale=0.9]
    \draw (0,0)node[font=\large]{$D_{\frac{n}{2}+1}$};
    \draw[<->,>=stealth] (3.75,-0.0) to node[below,font=\small]{bosonize} node[above,font=\small]{fermionize}(0.75,-0.0);
    \draw (4.5,0)node[font=\large]{$B_{\frac{n}{2}}$};
    \draw (0,3)node[font=\large]{$A_{n-1}$};
    \draw[<->,>=stealth] (0.75,3) to node[below,font=\small]{bosonize} node[above,font=\small]{fermionize} (3.75,3.0);
    \draw (4.5,3) node[font=\large]{${C_{\frac{n}{2}}}$};
    \draw[<->,>=stealth] (4.5,0.5) to node[right,font=\small]{stack} (4.5,2.5);
    \draw[<->,>=stealth] (0,0.5) to node[left,font=\small]{orbifold} (0,2.5);
    \end{tikzpicture}
    \end{minipage}
    \begin{minipage}{0.4\textwidth}
    \centering
    \begin{tikzpicture}[transform shape, scale=0.9]
    \draw (0,0)node[font=\large]{$E_6$};
    \draw[<->,>=stealth] (3.5,-0.0) to node[below,font=\small]{bosonize} node[above,font=\small]{fermionize}(0.5,-0.0);
    \draw (4,0)node[font=\large]{$F_4$};
    \draw[->,>=stealth] (-0.4,0.3) arc (225:-45:0.45);
    \draw[->,>=stealth] (3.6,0.3) arc (225:-45:0.45);
    \draw (-0.2,1.4) node[font=\small] {orbifold};
    \draw (3.9,1.4) node[font=\small] {stack};
    \end{tikzpicture}
    \end{minipage}
    \caption{The relationship between bosonic modular invariants and their fermionizations. The left panel shows the bosonization/fermionization maps of modular invariants $A_{n-1}$ and $D_{\frac{n}{2}+1}$ for an even $n$. The right panel shows the maps associated with modular invariant $E_6$.}
    \label{fig:su2_ferm}
\end{figure}

\subsection{Fermionic model from \texorpdfstring{$E_6$}{E6}}
Although exceptional modular invariants $E_7$ and $E_8$ do not have any global symmetry, the modular invariant $E_6$ at level $k=10$ has a $\BZ_2$ symmetry without anomaly~\cite{Lienart:2000jw}.
Using this, we can fermionize the exceptional modular invariant $E_6$.
The $\BZ_2$ twisted partition functions of $E_6$ invariant are given by
\begin{align}
\begin{aligned}
    Z_{E_6}[0,0] &= |\chi_1+\chi_7|^2 + |\chi_4+\chi_8|^2 +|\chi_5+\chi_{11}|^2\,,\\
    Z_{E_6}[0,1] &= |\chi_1+\chi_7|^2 - |\chi_4+\chi_8|^2 +|\chi_5+\chi_{11}|^2\,,\\
    Z_{E_6}[1,0] &= |\chi_4+\chi_8|^2 + (\chi_1+\chi_7)(\bar{\chi}_5+\bar{\chi}_{11}) +(\chi_5+\chi_{11})(\bar{\chi}_1 + \bar{\chi}_7)\,,\\
    Z_{E_6}[1,1] &= |\chi_4+\chi_8|^2 - (\chi_1+\chi_7)(\bar{\chi}_5+\bar{\chi}_{11}) -(\chi_5+\chi_{11})(\bar{\chi}_1 + \bar{\chi}_7)\,.
\end{aligned}
\end{align}
The orbifold of the $\BZ_2$ symmetry reproduces the modular invariant $E_6$ and the theory is self-dual under $\BZ_2$ gauging.
The fermionized theory has the torus partition functions
\begin{align}
\label{eq:f4partition}
    \begin{aligned}
        Z_{\CF,\,E_6}[0,0] &= |\chi_1+\chi_5+\chi_7+\chi_{11}|^2\,, &
        Z_{\CF,\,E_6}[0,1] &= |\chi_1-\chi_5+\chi_7-\chi_{11}|^2\,,\\
        Z_{\CF,\,E_6}[1,0] &= 2\,|\chi_4+\chi_8|^2 \,, & Z_{\CF,\,E_6}[1,1] &= 0\,.
    \end{aligned}
\end{align}
We can see that the fermionic partition functions are block-diagonal.
The diagonal terms $|\chi_p|^2$ consist of $p=1,5,7,11$.
This suggests a correspondence to the Dynkin diagram $F_4$ because its exponents are $m=1,5,7,11$.
Also, the height $n=12$ is exactly the Coxeter number of the Dynkin diagram $F_4$. See table~\ref{tab:coxeter_extended}.
Thus, the fermionized theory of modular invariant $E_6$ is called $F_4$.\footnote{A partition function related to Dynkin diagram $F_4$ were discussed in~\cite{Zuber:1993vm}. Their construction was the projection onto the $\BZ_2$-even sector in the original modular invariant $E_6$, while we further add the twisted sector to the spectrum for a theory to be consistent by itself.}

The stacking of the Kitaev-Majorana chain does not affect the partition functions due to the vanishing R-R partition function. This signals self-duality under gauging the $\BZ_2$ symmetry in the modular invariant $E_6$ (see the right of Fig.~\ref{fig:su2_ferm}). We remark that the folding of the Dynkin diagram $E_6$ yields the Dynkin diagram $F_4$, which supports the relationship between fermionization and folding of the Dynkin diagram.

Note that the fermionized theory $F_4$ can be understood from the conformal embedding $\widehat{su}(2)_{10}\subset \widehat{sp}(4)_1$.
The affine $\widehat{sp}(4)_1$ algebra has three primaries labeled by $(0,0)$, $(0,1)$, and $(1,0)$, whose conformal dimensions are $h_{(0,0)}=0$, $h_{(0,1)}=1/2$, and $h_{(1,0)}=5/16$.
The characters of the two algebras are related by (refer to section 17.5.2 of~\cite{DiFrancesco:1997nk})
\begin{align}
    \chi_{(0,0)} = \chi_1+\chi_7\,,\qquad  \chi_{(0,1)} = \chi_5+\chi_{11}\,,\qquad
    \chi_{(1,0)} = \chi_4+\chi_8\,.
\end{align}
Thus, we can rewrite the torus partition functions~\eqref{eq:f4partition} in terms of the affine $\widehat{sp}(4)_1$ characters.

To diagnose the profile of the fermionic theory $F_4$, consider the small $q$ expansion of the NS-NS partition function
\begin{align}
    Z_{\CF,\,E_6}[0,0] = q^{-\frac{5}{48}}\,\bar{q}^{-\frac{5}{48}} \left(1+5\,q^{\frac{1}{2}} + 5\,\bar{q}^{\frac{1}{2}} + 10\,q + 25\,q^{\frac{1}{2}}\,\bar{q}^{\frac{1}{2}} + 10\,\bar{q}+\cdots  \right)\,.
\end{align}
This partition function contains the terms $q^{1/2}$ and $\bar{q}^{1/2}$, which reflects the inclusion of free fermions.
In fact, this exactly agrees with the NS-NS partition function of $5$ copies of free Majorana fermion with central charge $(c,\bar{c})=(\frac{1}{2},\frac{1}{2})$
\begin{align}
    Z_{\psi^{\otimes5}}[0,0] = \left|\frac{\theta_3(q)}{\eta(q)}\right|^{5}\,,
\end{align}
where $\theta_3(q):=\sum_{n\in\BZ}q^{\frac{n^2}{2}}$ is the Jacobi theta function and $\eta(q)$ is the Dedekind eta function.
Thus, the fermionization of $\widehat{su}(2)_{10}$ model with modular invariant $E_6$ can be described by $5$ copies of Majorana fermions.

\subsection{Parafermionic models from \texorpdfstring{$D_4$}{D4}}

Let us move on to the classification of parafermionic $\widehat{su}(2)$ WZW models.
The parafermionization is applicable when the theory has a non-anomalous $\BZ_{N\geq3}$ symmetry.
For $\widehat{su}(2)_k$ models, such a symmetry arises only in the theory with modular invariant $D_4$ at level $k=4$.
The non-anomalous symmetry group of the corresponding theory is the permutation group $S_3$~\cite{Lienart:2000jw}.
In addition to the $\BZ_2$ symmetry we fermionized in \eqref{eq:btype-odd}, the group $S_3$ contains a cyclic group $\BZ_3$ as a subgroup.
The torus partition functions twisted by the $\BZ_3$ symmetry are
\begin{align}
\begin{aligned}
    Z_{D_4}[0,0] &= |\chi_1+\chi_5|^2 +2\,|\chi_3|^2\,,\qquad 
    Z_{D_4}[0,1] = Z_{D_4}[0,2] = |\chi_1+\chi_5|^2 -|\chi_3|^2\,,\\
    Z_{D_4}[1,a] &= |\chi_3|^2 + \omega_3^a\,\chi_3\,(\bar{\chi}_1+\bar{\chi}_5) + \omega_3^{2a}\,(\chi_1+\chi_5)\,\bar{\chi}_3\,,\\
    Z_{D_4}[2,a] &=  |\chi_3|^2 + \omega_3^{2a}\,\chi_3\,(\bar{\chi}_1+\bar{\chi}_5) + \omega_3^{a}\,(\chi_1+\chi_5)\,\bar{\chi}_3\,,
\end{aligned}
\end{align}
where $\omega_3 = \exp(2\pi\i/3)$.
The orbifold of modular invariant $D_4$ by the $\BZ_3$ symmetry returns back to itself.
The $\BZ_3$ parafermionization with $\rho=1$ gives the following partition functions:
\begin{align}
\begin{aligned}
\label{eq:paraf_d4}
    Z_{\mathrm{PF}_1,\,D_4}[0,0] &= |\chi_1+\chi_5|^2 +2\,(\chi_1+{\chi}_5) \,\bar{\chi}_3\,,\\
    Z_{\mathrm{PF}_1,\,D_4}[0,1] &= Z_{\mathrm{PF}_1,\,D_4}[0,2] = |\chi_1+\chi_5|^2 - (\chi_1+{\chi}_5) \,\bar{\chi}_3\,,\\
    Z_{\mathrm{PF}_1,\,D_4}[1,k] &= \omega_3^{2k}\left(\chi_3\,(\bar{\chi}_1+\bar{\chi}_5) + (\omega_3^k+\omega_3^{2k})\,\chi_3\,\bar{\chi}_3\right)\,,\\
    Z_{\mathrm{PF}_1,\,D_4}[2,k] &= \omega_3^{k}\left(\chi_3\,(\bar{\chi}_1+\bar{\chi}_5) + (\omega_3^k+\omega_3^{2k})\,\chi_3\,\bar{\chi}_3\right)\,.
\end{aligned}
\end{align}
We can easily check that the partition functions are invariant under stacking and conjugation given by~\eqref{eq:relpara_1}, which is consistent with the self-duality under the orbifold.
Notably, we can see that the parafermionic partition function $Z_{\mathrm{PF}_{1},\,D_4}[0,0]$ contains the diagonal terms $|\chi_p|^2$ with $p=1,5$.
This suggests a relation to the Dynkin diagram $G_2$ since its exponents are just 1 and 5.
Furthermore, the height $n=6$ is the Coxeter number of the Dynkin diagram $G_2$.
We denote by $G_2$ the parafermionic theory with the partition functions~\eqref{eq:paraf_d4}.

Let us see the spectrum of the parafermionic theory $G_2$.
The affine $\widehat{su}(2)_4$ primary operators labeled by $p=1,3,5$ have conformal dimensions $h=0,1/3,1$, respectively.
Since the partition function with paraspin structure $(k_1,k_2) = (0,0)$ contains the terms $\chi_1\bar{\chi}_3$ and $\chi_5\bar{\chi}_3$, there exists fractional spin operators with spin proportional to $1/3$ in its spectrum.
Similarly, from the partition functions with paraspin structure $(k_1,k_2) = (1,0),(2,0)$, one can see that the other sectors also include a fractional spin.

We summarize the relationship between $D_4$ and its parafermionization in Fig.~\ref{fig:paraf_d4}.
The original bosonic theory $D_4$ is self-dual under $\BZ_3$ orbifolding and the parafermionized theory $G_2$ is self-dual under stacking and conjugation.
Notably, the folding of the Dynkin diagram $D_4$ by the $\BZ_3$ automorphism yields the Dynkin diagram $G_2$.
Thus, the $\BZ_3$ parafermionization can be interpreted as folding by $\BZ_3$ automorphism in terms of the associated Dynkin diagram.

There is another parafermionization by choosing $\rho=2$.
The corresponding partition functions are given by
\begin{align}
\begin{aligned}
\label{eq:paraf2_d4}
    Z_{\mathrm{PF}_2,\,D_4}[0,0] &= |\chi_1+\chi_5|^2 +2\, \chi_3\,(\bar{\chi}_1+\bar{\chi}_5)\,,\\
    Z_{\mathrm{PF}_2,\,D_4}[0,1] &= Z_{\mathrm{PF}_2,\,D_4}[0,2] = |\chi_1+\chi_5|^2 - \chi_3\,(\bar{\chi}_1+\bar{\chi}_5)\,,\\
    Z_{\mathrm{PF}_2,\,D_4}[1,k] &= \omega_3^{k}\left((\chi_1+\chi_5)\,\bar{\chi}_3 + (\omega_3^k+\omega_3^{2k})\,\chi_3\,\bar{\chi}_3\right)\,,\\
    Z_{\mathrm{PF}_2,\,D_4}[2,k] &= \omega_3^{2k}\left((\chi_1+\chi_5)\,\bar{\chi}_3 + (\omega_3^k+\omega_3^{2k})\,\chi_3\,\bar{\chi}_3\right)\,.
\end{aligned}
\end{align}
This theory is also invariant under stacking and conjugation as in the case of $\rho=1$. 
Note that these parafermionic partition functions are complex conjugates of \eqref{eq:paraf_d4} with $\rho=1$.
We call this parafermionic theory by $\overline{G_2}$.
As in $\rho =1$, the parafermionized theory with $\rho=2$ includes a fractional spin operator, while the holomorphic and anti-holomorphic parts are flipped.

\begin{figure}
    \centering
    \begin{tikzpicture}[transform shape, scale=0.9]
    \draw (0,0)node[font=\large]{$D_4$};
    \draw[<->,>=stealth] (3.5,-0.0) to node[below,font=\small]{parafermionize} node[above,font=\small]{bosonize}(0.5,-0.0);
    \draw (4,0)node[font=\large]{$G_2$};
    \draw[->,>=stealth] (-0.4,0.3) arc (225:-45:0.45);
    \draw[->,>=stealth] (3.6,0.3) arc (225:-45:0.45);
    \draw (-0.2,1.4) node[font=\small] {orbifold};
    \draw (3.8,1.1) node[above,font=\small,align=center] {stack\\+conjugate};
    \end{tikzpicture}
    \caption{Parafermionization of modular invariant $D_4$.}
    \label{fig:paraf_d4}
\end{figure}

\section{Classification of \texorpdfstring{$\widehat{su}(3)$}{su(3)} models}
\label{sec:su(3)}

This section is devoted to classifying fermionic and parafermionic $\widehat{su}(3)$ models using fermionization and parafermionization.
First, we summarize the classification result of fermionic and parafermionic $\widehat{su}(3)$ models.
We show the list of fermionic and parafermionic theories with the affine $\widehat{su}(3)$ algebra in table~\ref{tab:listsu(3)}.
The list provides the height and the torus partition functions for each fermionic and parafermionic theory.

In table~\ref{tab:listsu(3)}, there are seven theories (the above six are parafermionic and the last one is fermionic). Among them, the above two are infinite series at height $n\geq 4$. The other five theories are exceptional and only appear at the special height $n=8,12,6$.
The six parafermionic theories form three pairs for $n\geq4$, $n=8$, and $n=12$, which correspond to the parafermionization of $\CA_n$, $\CE_8$, and $\CE_{12}$, respectively (see Fig.~\ref{fig:atypedtypesu(3)} and Fig.~\ref{fig:para-ce8}). Each pair consists of parafermionization with $\rho=1$ and $\rho=2$.
The only fermionic theory is the fermionization of modular invariant $\CD_6$ (see the right of Fig.~\ref{fig:atypedtypesu(3)}).
The fermionic theory is self-dual under the stacking of the Arf theory.
Note that the list classifies parafermionic CFTs up to charge-conjugation and the operation~\eqref{eq:para_orb} consisting of $\mathrm{Arf}_N$ stacking and conjugation.
The last three theories at height $n=12$ and $n=6$ are self-dual under charge conjugation.

\begin{table}[tb]
    \centering
    \begin{tabular*}{\textwidth}{c@{\hskip 0.8cm}l}
    \toprule\addlinespace

    \textbf{Height} & \textbf{Fermionic or parafermionic partition functions} \\
    \addlinespace\midrule\addlinespace[0.4cm]

    $n\geq 4$ & \small{$\displaystyle \sum_{t(p)=0} |\chi_p|^2 + \sum_{t(\mu(p))=2} \chi_{\mu^2(p)}\,\bar{\chi}_p +  \sum_{t(\mu^2(p))=1} \chi_{\mu(p)}\,\bar{\chi}_p$} \\[0.8cm]

    $n\geq 4$ & \small{$\displaystyle \sum_{t(p)=0} |\chi_p|^2 +  \sum_{t(\mu^2(p))=2} \chi_{\mu(p)}\,\bar{\chi}_p +  \sum_{t(\mu(p))=1} \chi_{\mu^2(p)}\,\bar{\chi}_p$}\, \\[0.8cm]

    $n=8$ & \makecell[{{l}}]{
    \small{$\displaystyle \left(\chi_{(1,1)} +\chi_{(3,3)}+ \chi_{(1,6)}+\chi_{(3,2)}+\chi_{(2,3)}+\chi_{(6,1)}\right)\left(\bar{\chi}_{(1,1)}+\bar{\chi}_{(3,3)}\right)$} \\ \small{$+\left(\chi_{(1,4)}+\chi_{(4,1)}+\chi_{(1,3)} + \chi_{(4,3)}+\chi_{(3,1)} + \chi_{(3,4)}\right)\left(\bar{\chi}_{(1,4)}+\bar{\chi}_{(4,1)}\right)$}} \\[0.8cm]

    $n=8$ & \makecell[{{l}}]{
    \small{$\displaystyle \left({\chi}_{(1,1)}+{\chi}_{(3,3)}\right)\left(\bar{\chi}_{(1,1)} +\bar{\chi}_{(3,3)}+\bar{\chi}_{(1,6)}+\bar{\chi}_{(3,2)}+\bar{\chi}_{(2,3)}+\bar{\chi}_{(6,1)})\right)$} \\
    \small{$+ \left({\chi}_{(1,4)}+{\chi}_{(4,1)}\right)\left(\bar{\chi}_{(1,4)}+\bar{\chi}_{(4,1)}+\bar{\chi}_{(1,3)} + \bar{\chi}_{(4,3)}+\bar{\chi}_{(3,1)} + \bar{\chi}_{(3,4)}\right)$
    }}\\[0.8cm]

    $n=12$ & \makecell[{{l}}]{
    \small{$\displaystyle \left(\chi_{(1,1)}+\chi_{(1,10)}+\chi_{(10,1)}+\chi_{(2,5)}+\chi_{(5,2)}+\chi_{(5,5)}+2\left(\chi_{(3,3)}+\chi_{(3,6)}+\chi_{(6,3)}\right)\right)$}
    \\
    \small{$\times \left(\bar{\chi}_{(1,1)}+\bar{\chi}_{(1,10)}+\bar{\chi}_{(10,1)}+\bar{\chi}_{(2,5)}+\bar{\chi}_{(5,2)}+\bar{\chi}_{(5,5)}\right)$}}\\[0.8cm]

    $n=12$ & \makecell[{{l}}]{
    \small{$\displaystyle \left(\chi_{(1,1)}+\chi_{(1,10)}+\chi_{(10,1)}+\chi_{(2,5)}+\chi_{(5,2)}+\chi_{(5,5)}\right)$}\\
    \small{$\times\left(\bar{\chi}_{(1,1)}+\bar{\chi}_{(1,10)}+\bar{\chi}_{(10,1)}+\bar{\chi}_{(2,5)}+\bar{\chi}_{(5,2)}+\bar{\chi}_{(5,5)}+ 2\left(\bar{\chi}_{(3,3)}+\bar{\chi}_{(3,6)}+\bar{\chi}_{(6,3)}\right)\right)$}}\\[0.5cm]
    
    \addlinespace\midrule\addlinespace[0.2cm]

    $n=6$ & \small{$\displaystyle Z_{\CF}[0,0] = |\chi_{(1,1)} + \chi_{(1,4)} + \chi_{(4,1)}+\chi_{(2,2)}|^2\,\qquad Z_{\CF}[1,1]=0$}\\ \addlinespace[0.2cm]
    
    \bottomrule    
    \end{tabular*}
    \caption{List of fermionic and parafermionic CFTs with affine $\widehat{su}(3)$ algebra symmetry at each height $n=k+3$. 
    For the six types of the parafermionic theory above, we show the partition function $Z_\mathrm{PF}[0,0]$ on the paraspin structure $(k_1,k_2) =(0,0)$.
    For the last fermionic theory with $n=6$, we show the NS-NS and R-R partition functions.
    We omit the theories related to the above list by the charge conjugation and the operation~\eqref{eq:para_orb} consisting of the $\mathrm{Arf}_N$ stacking and conjugation.}
    \label{tab:listsu(3)}
\end{table}

\subsection{Parafemionic models from \texorpdfstring{$\CA_n^{(\ast)}$}{An} and \texorpdfstring{$\CD_n^{(\ast)}$}{Dn}}

First, we consider the parafermionization of the diagonal modular invariants $\CA_n$.
The bosonic theories always have a non-anomalous $\BZ_3$ symmetry.
The twisted partition functions are
\begin{align}
\begin{aligned}
\label{eq:twistpart_An}
    Z_{\CA_n}[0,a] &= \sum_p\,\omega_3^{at(p)}\,|\chi_p|^2\,,\\
    Z_{\CA_n}[1,a] &= \sum_p \omega_3^{at(\mu(p))}\,\chi_{\mu^2(p)}\,\bar{\chi}_p\,,\\
    Z_{\CA_n}[2,a] &= \sum_p \omega_3^{at(\mu^2(p))}\,\chi_{\mu(p)}\,\bar{\chi}_p\,.
\end{aligned}
\end{align}
As mentioned in section~\ref{sec:modularinv}, the triality $t(p)$ is invariant under automorphism $\mu$ when $n\in3\BZ$, but otherwise it is shifted to $t(\mu(p))=t(p)+n$ mod $3$.
After orbifolding by the $\BZ_3$ symmetry, we obtain the modular invariant $\CD_\ell$ $(\ell\notin3\BZ)$ when $n\notin3\BZ$ and $\CD_\ell$ $(\ell\in3\BZ)$ when $n\in3\BZ$.

Associated with the $\BZ_3$ symmetry, we can parafermionize the diagonal modular invariant $\CA_n$.
The parafermionized theory of the diagonal modular invariant $\CA_n$ with $\rho=1$ has the partition functions
\begin{align}
    \begin{aligned}
    \label{eq:para_ca1}
        Z_{\mathrm{PF}_1,\,\CA_n}[0,k] &= \sum_{t(p)=0} |\chi_p|^2 + \omega_3^k \sum_{t(\mu(p))=2} \chi_{\mu^2(p)}\,\bar{\chi}_p + \omega_3^{2k} \sum_{t(\mu^2(p))=1} \chi_{\mu(p)}\,\bar{\chi}_p \,,\\
        Z_{\mathrm{PF}_1,\,\CA_n}[1,k] &= \sum_{t(\mu^2(p))=0} \chi_{\mu(p)}\,\bar{\chi}_p + \omega_3^k \sum_{t(p)=2} |\chi_p|^2 + \omega_3^{2k} \sum_{t(\mu(p))=1} \chi_{\mu^2(p)}\,\bar{\chi}_p \,,\\
        Z_{\mathrm{PF}_1,\,\CA_n}[2,k] &= \sum_{t(\mu(p))=0} \chi_{\mu^2(p)}\,\bar{\chi}_p + \omega_3^k \sum_{t(\mu^2(p))=2} \chi_{\mu(p)}\,\bar{\chi}_p + \omega_3^{2k} \sum_{t(p)=1} |\chi_p|^2 \,.
    \end{aligned}
\end{align}
On the other hand, the parafermionized theory with $\rho=2$ is given by
\begin{align}
    \begin{aligned}
    \label{eq:para_ca2}
        Z_{\mathrm{PF}_2,\,\CA_n}[0,k] &= \sum_{t(p)=0} |\chi_p|^2 + \omega_3^k \sum_{t(\mu^2(p))=2} \chi_{\mu(p)}\,\bar{\chi}_p + \omega_3^{2k} \sum_{t(\mu(p))=1} \chi_{\mu^2(p)}\,\bar{\chi}_p \,,\\
        Z_{\mathrm{PF}_2,\,\CA_n}[1,k] &= \sum_{t(\mu^2(p))=0} \chi_{\mu(p)}\,\bar{\chi}_p + \omega_3^k \sum_{t(\mu(p))=2} \chi_{\mu^2(p)}\,\bar{\chi}_p + \omega_3^{2k} \sum_{t(p)=1} |\chi_p|^2 \,,\\
        Z_{\mathrm{PF}_2,\,\CA_n}[2,k] &= \sum_{t(\mu(p))=0} \chi_{\mu^2(p)}\,\bar{\chi}_p + \omega_3^k \sum_{t(p)=2} |\chi_p|^2 + \omega_3^{2k} \sum_{t(\mu^2(p))=1} \chi_{\mu(p)}\,\bar{\chi}_p \,.
    \end{aligned}
\end{align}
We can also start with its orbifold, i.e., the modular invariant $\CD_n$.
The modular invariant $\CD_n$ has a non-anomalous $\BZ_3$ symmetry, which is the dual symmetry of the diagonal modular invariant $\CA_n$. 
From \eqref{eq:gene_orb}, the twisted partition functions of $\CD_n$ are
\begin{align}
\label{eq:twistpart_Dn}
    Z_{\CD_n}[a_1,a_2] = \sum_{j=0}^2 \,\sum_{t(\mu^j(p))=2a_1} \omega_3^{2ja_2}\,\chi_{\mu^{2j}(p)}\,\bar{\chi}_p\,.
\end{align}
We can parafermionize the modular invariant $\CD_n$ using the $\BZ_3$ symmetry.
The parafermionized partition functions with $\rho=1$ are
\begin{align}
    \begin{aligned}
        Z_{\mathrm{PF}_1,\,\CD_n}[0,k] &= \sum_{t(p)=0} |\chi_p|^2 + \omega_3^k \sum_{t(\mu(p))=2} \chi_{\mu^2(p)}\,\bar{\chi}_p + \omega_3^{2k} \sum_{t(\mu^2(p))=1} \chi_{\mu(p)}\,\bar{\chi}_p \,,\\
        Z_{\mathrm{PF}_1,\,\CD_n}[1,k] &= \sum_{t(p)=1} |\chi_p|^2 + \omega_3^{k} \sum_{t(\mu(p))=0} \chi_{\mu^2(p)}\,\bar{\chi}_p + \omega_3^{2k} \sum_{t(\mu^2(p))=2} \chi_{\mu(p)}\,\bar{\chi}_p\,,\\
        Z_{\mathrm{PF}_1,\,\CD_n}[2,k] &= \sum_{t(p)=2} |\chi_p|^2 + \omega_3^{k} \sum_{t(\mu(p))=1} \chi_{\mu^2(p)}\,\bar{\chi}_p + \omega_3^{2k} \sum_{t(\mu^2(p))=0} \chi_{\mu(p)}\,\bar{\chi}_p\,.
    \end{aligned}
\end{align}
These partition functions are related to the parafermionized theory~\eqref{eq:para_ca1} of the diagonal modular invariant $\CA_n$ by~\eqref{eq:relpara_1}.
The parafermionized partition functions with $\rho=2$ are
\begin{align}
    \begin{aligned}
        Z_{\mathrm{PF}_2,\,\CD_n}[0,k] &= \sum_{t(p)=0} |\chi_p|^2 + \omega_3^k \sum_{t(\mu(p))=1} \chi_{\mu^2(p)}\,\bar{\chi}_p + \omega_3^{2k} \sum_{t(\mu^2(p))=2} \chi_{\mu(p)}\,\bar{\chi}_p \,,\\
        Z_{\mathrm{PF}_1,\,\CD_n}[1,k] &= \sum_{t(p)=1} |\chi_p|^2 + \omega_3^{k} \sum_{t(\mu(p))=2} \chi_{\mu^2(p)}\,\bar{\chi}_p + \omega_3^{2k} \sum_{t(\mu^2(p))=0} \chi_{\mu(p)}\,\bar{\chi}_p\,,\\
        Z_{\mathrm{PF}_1,\,\CD_n}[2,k] &= \sum_{t(p)=2} |\chi_p|^2 + \omega_3^{k} \sum_{t(\mu(p))=0} \chi_{\mu^2(p)}\,\bar{\chi}_p + \omega_3^{2k} \sum_{t(\mu^2(p))=1} \chi_{\mu(p)}\,\bar{\chi}_p\,.
    \end{aligned}
\end{align}
These partition functions are related to the parafermionized theory~\eqref{eq:para_ca2} of the diagonal modular invariant $\CA_n$ by~\eqref{eq:relpara_2}.
We summarize the relationship among the bosonic theories $\CA_n$, $\CD_n$, and their parafermionization with a fixed parameter $\rho\in\{1,2\}$ in the left panel of Fig.~\ref{fig:atypedtypesu(3)}. 
One can start with the charge-conjugated theories $\CA_n^\ast$ and $\CD_n^\ast$.
Since those theories have a $\BZ_3$ symmetry and their twisted partition functions are exactly charge conjugate of \eqref{eq:twistpart_An} and \eqref{eq:twistpart_Dn}, the parafermionized theories of $\CA_n^\ast$ and $\CD_n^\ast$ are given by charge conjugate of the above parafermionic partition functions.

\begin{figure}
    \centering
    \begin{minipage}{0.48\textwidth}
    \begin{tikzpicture}[transform shape, scale=0.9]
    \draw (0,0)node[font=\large]{$\CD_n$};
    \draw[<->,>=stealth] (3.75,-0.0) to node[below,font=\small]{bosonize} node[above,font=\small]{parafermionize}(0.75,-0.0);
    \draw (4.5,0)node[font=\large]{$\mathrm{PF}[\CD_n]$};
    \draw (0,3)node[font=\large]{$\CA_n$};
    \draw[<->,>=stealth] (0.75,3) to node[below,font=\small]{bosonize} node[above,font=\small]{parafermionize} (3.75,3.0);
    \draw (4.5,3) node[font=\large]{$\mathrm{PF}[\CA_n]$};
    \draw[<->,>=stealth] (4.5,0.5) to node[right,font=\small,align=left]{stack\\+conjugate} (4.5,2.5);
    \draw[<->,>=stealth] (0,0.5) to node[left,font=\small]{orbifold} (0,2.5);
    \end{tikzpicture}
    \end{minipage}\quad
    \begin{minipage}{0.48\textwidth}
    \begin{tikzpicture}[transform shape, scale=0.9]
    \draw (0,0)node[font=\large]{$\CD_{6}$};
    \draw[<->,>=stealth] (3.5,-0.0) to node[below,font=\small]{fermionize} node[above,font=\small]{bosonize}(0.5,-0.0);
    \draw (4.2,0)node[font=\large]{$\CF[\CD_{6}]$};
    \draw[->,>=stealth] (-0.4,0.3) arc (225:-45:0.45);
    \draw[->,>=stealth] (3.8,0.3) arc (225:-45:0.45);
    \draw (-0.2,1.4) node[font=\small] {orbifold};
    \draw (4.1,1.4) node[font=\small] {stack};
    \end{tikzpicture}
    \end{minipage}
    \caption{The parafermionization of modular invariants $\CA_n$ and $\CD_n$ with a given $\rho$ (the left), and the fermionization of modular invariant $\CD_6$ (the right). For the theory $\CT$, we denote its fermionization by $\CF[\CT]$ and its parafermionization by $\mathrm{PF}[\CT]$.}
    \label{fig:atypedtypesu(3)}
\end{figure}

\subsection{Fermionic model from \texorpdfstring{$\CD_6$}{D6}}
Consider the $\CD$-type modular invariant $\CD_6$ at level $3$:
\begin{align}
\label{eq:cd6}
    Z_{\CD_6}[0,0] = |\chi_{(1,1)}+\chi_{(1,4)}+\chi_{(4,1)}|^2 + 3\,|\chi_{(2,2)}|^2\,,
\end{align}
where this bosonic theory is invariant under charge conjugation: $\CD_6=\CD_6^\ast$.
The global symmetry of this theory is known to be the alternating group $A_4$ of degree $4$~\cite{Lienart:2000jw}, which contains one $\BZ_3$ group and one $\BZ_2\times\BZ_2$ group as a maximal abelian subgroup.
Let us denote generators of the $\BZ_2\times \BZ_2$ symmetry by $g$ and $g'$: $\BZ_2\times \BZ_2 = \langle g, g'\rangle$.
Then, the modular invariant $\CD_6$ admits the action of three $\BZ_2$ subgroups $\langle g\rangle,\,\langle g'\rangle,\,\langle gg'\rangle$.
However, the three $\BZ_2$ symmetries act on the partition function \eqref{eq:cd6} in the same way. More precisely, they act on the three degenerate fields $|\chi_{(2,2)}|^2$ by $g=\mathrm{diag} (1,-1,-1)$, $g'=\mathrm{diag} (-1,1,-1)$, $gg'=\mathrm{diag} (-1,-1,1)$.
Thus, the twisted partition functions are the same and can be written as
\begin{align}
    \begin{aligned}
        Z_{\CD_6}[0,1] &= |\chi_{(1,1)}+\chi_{(1,4)}+\chi_{(4,1)}|^2 -|\chi_{(2,2)}|^2\,,\\
        
        Z_{\CD_6}[1,0] &= (\chi_{(2,2)}+\chi_{(1,1)}+\chi_{(1,4)}+\chi_{(4,1)})\,\bar{\chi}_{(2,2)} + \chi_{(2,2)} \,(\bar{\chi}_{(2,2)}+\bar{\chi}_{(1,1)}+\bar{\chi}_{(1,4)}+\bar{\chi}_{(4,1)})\,,\\
        
        Z_{\CD_6}[1,1] &= (\chi_{(2,2)}-\chi_{(1,1)}-\chi_{(1,4)}-\chi_{(4,1)})\,\bar{\chi}_{(2,2)} + \chi_{(2,2)} \,(\bar{\chi}_{(2,2)}-\bar{\chi}_{(1,1)}-\bar{\chi}_{(1,4)}-\bar{\chi}_{(4,1)})\,.
    \end{aligned}
\end{align}
Note that the $\BZ_2$ twisted partition functions are still invariant under charge conjugation, inherited from the original modular invariant $\CD_6=\CD_6^\ast$.
After orbifolding by the $\BZ_2$ symmetry, the partition function is again given by
\begin{align}
    Z_{\CO,\,\CD_6}[0,0] = |\chi_{(1,1)}+\chi_{(1,4)}+\chi_{(4,1)}|^2 + 3\,|\chi_{(2,2)}|^2\,.
\end{align}
Thus, the modular invariant $\CD_6$ is self-dual under the $\BZ_2$ orbifold.

As mentioned above, the three non-anomalous $\BZ_2$ symmetries in the bosonic theory $\CD_6$ yield the same twisted partition functions.
As a consequence, the fermionized partition functions are also irrelevant to which $\BZ_2$ symmetry is gauged and are given by
\begin{align}
\begin{aligned}
\label{eq:fermcd6}
    Z_{\CF,\,\CD_6}[0,0] &= |\chi_{(1,1)} + \chi_{(1,4)} + \chi_{(4,1)}+\chi_{(2,2)}|^2\,,\\
    Z_{\CF,\,\CD_6}[0,1] &= |\chi_{(1,1)} + \chi_{(1,4)} + \chi_{(4,1)}-\chi_{(2,2)}|^2\,,\\
    Z_{\CF,\,\CD_6}[1,0] &= 4\,|\chi_{(2,2)}|^2\,,\qquad Z_{\CF,\,\CD_6}[1,1]=0\,.
\end{aligned}
\end{align}
Since the $\BZ_2$ twisted partition functions are invariant under charge conjugation $C$, the fermionized theory is also self-conjugate. 
Since the R-R partition function is zero, this fermionic theory is invariant under the stacking of the Arf theory. This is consistent with the self-duality of $\CD_6$ under orbifolding.
We summarize the relationship between the bosonic theory $\CD_6$ and the fermionized theory in the right of Fig.~\ref{fig:atypedtypesu(3)}.

To capture the profiles of the fermionic theory,
we use the conformal embedding $\widehat{su}(3)_3\subset \widehat{so}(8)_1$~\cite{Schellekens:1986mb,Bais:1986zs} to rewrite the partition functions in terms of $\widehat{so}(8)_1$ characters.
Since $\widehat{so}(8)_1$ has four primaries labeled by $0$, $v$, $s$, and $c$, whose conformal dimensions are $h_0 = 0$ and $h_v=h_s=h_c=1/2$, respectively.
The characters of $\widehat{su}(3)_3$ and $\widehat{so}(8)_1$ are related by (\cite{Christe:1988vc})
\begin{align}
    \chi_0 = \chi_{(1,1)} + \chi_{(1,4)} + \chi_{(4,1)} 
    \,,\qquad \chi_v = \chi_s = \chi_c= \chi_{(2,2)}\,,
\end{align}
where the second equation implies the triality of the three non-trivial representations $v,s,c$ in $\widehat{so}(8)_1$ algebra.
From these relations, we can represent the fermionic partition functions~\eqref{eq:fermcd6} by $\widehat{so}(8)_1$ characters, which can be described by four pairs of Dirac fermions.

\subsection{Parafermionic models from \texorpdfstring{$\CE_8^{(\ast)}$}{E8}}
Let us consider an exceptional modular invariant $\CE_8$ of $\widehat{su}(3)$ at level $k=5$.
This modular invariant is not self-conjugate: $\CE_8\not = \CE_8^\ast$.
The non-anomalous global symmetry is a cyclic group $\BZ_3$ and the corresponding twisted partition functions of $\CE_8$ are given by
\begin{align}
    \begin{aligned}
    \label{eq:twste8}
        Z_{\CE_8}[0,k]&= |\chi_{(1,1)}+\chi_{(3,3)}|^2+\omega_3^k\,|\chi_{(3,1)}+\chi_{(3,4)}|^2 + \omega_3^{2k}\,|\chi_{(1,3)}+\chi_{(4,3)}|^2\\
        &\quad +|\chi_{(4,1)}+\chi_{(1,4)}|^2 + \omega_3^k\,|\chi_{(2,3)}+\chi_{(6,1)}|^2+\omega_3^{2k}\,|\chi_{(3,2)}+\chi_{(1,6)}|^2 \,,\\
        
        Z_{\CE_8}[1,k]&= \left(\chi_{(1,6)}+\chi_{(3,2)}\right)\left(\bar{\chi}_{(2,3)}+\bar{\chi}_{(6,1)}\right) + \left(\chi_{(1,3)}+\chi_{(4,3)}\right)\left(\bar{\chi}_{(3,1)}+\bar{\chi}_{(3,4)}\right)\\ 
        &\quad + \omega_3^k \left[\left(\chi_{(1,1)}+\chi_{(3,3)}\right)\left(\bar{\chi}_{(1,6)}+\bar{\chi}_{(3,2)}\right)+ \left(\chi_{(1,4)}+\chi_{(4,1)}\right)\left(\bar{\chi}_{(1,3)}+\bar{\chi}_{(4,3)}\right)\right]\\
        &\quad + \omega_3^{2k} \left[\left(\chi_{(2,3)}+\chi_{(6,1)}\right)\left(\bar{\chi}_{(1,1)}+\bar{\chi}_{(3,3)}\right) + \left(\chi_{(3,1)}+\chi_{(3,4)}\right)\left(\bar{\chi}_{(1,4)}+\bar{\chi}_{(4,1)}\right)\right]\,,\\
        
        Z_{\CE_8}[2,k]&= \left(\chi_{(3,1)}+\chi_{(3,4)}\right)\left(\bar{\chi}_{(1,3)}+\bar{\chi}_{(4,3)}\right)  + \left(\chi_{(2,3)}+\chi_{(6,1)}\right)\left(\bar{\chi}_{(1,6)}+\bar{\chi}_{(3,2)}\right)\\ 
        &\quad + \omega_3^{2k}\left[\left(\chi_{(1,4)}+\chi_{(4,1)}\right)\left(\bar{\chi}_{(3,1)}+\bar{\chi}_{(3,4)}\right)+ \left(\chi_{(1,1)}+\chi_{(3,3)}\right)\left(\bar{\chi}_{(2,3)}+\bar{\chi}_{(6,1)}\right)\right]\\
        &\quad + \omega_3^k\left[\left(\chi_{(1,6)}+\chi_{(3,2)}\right)\left(\bar{\chi}_{(1,1)}+\bar{\chi}_{(3,3)}\right) + \left(\chi_{(1,3)}+\chi_{(4,3)}\right)\left(\bar{\chi}_{(1,4)}+\bar{\chi}_{(4,1)}\right)\right]\,.
    \end{aligned}
\end{align}
Note that the twisted partition functions of the conjugate theory $\CE_8^\ast$ are obtained by replacing the mass matrix $\CM_{p,p'}$ by $\CM_{p,C(p')}$. 
After $\BZ_3$ orbifold, the modular invariant $\CE_8$ turns into the charge-conjugated modular invariant $\CE_8^\ast$.
The parafermionized theory of $\CE_8$ with $\rho=1$ has the following partition functions:
\begin{align}
\begin{aligned}
\label{eq:parafe8}
    Z_{\mathrm{PF_1},\,\CE_8}[0,k] &= \left(\chi_{(1,1)} +\chi_{(3,3)}+\omega_3^{2k}\, (\chi_{(1,6)}+\chi_{(3,2)})+\omega_3^k\,(\chi_{(2,3)}+\chi_{(6,1)})\right)\left(\bar{\chi}_{(1,1)}+\bar{\chi}_{(3,3)}\right) \\
    &\quad+ \left(\chi_{(1,4)}+\chi_{(4,1)}+\omega_3^{2k}\,(\chi_{(1,3)} + \chi_{(4,3)})+\omega_3^k\,(\chi_{(3,1)} + \chi_{(3,4)})
    \right)\left(\bar{\chi}_{(1,4)}+\bar{\chi}_{(4,1)}\right)\,,\\

    Z_{\mathrm{PF_1},\,\CE_8}[1,k] &= \left(\chi_{(2,3)}+\chi_{(6,1)} +\omega_3^{2k}\,(\chi_{(1,1)}+\chi_{(3,3)})  + \omega_3^k\,(\chi_{(1,6)}+ \chi_{(3,2)})\right)\left(\bar{\chi}_{(1,6)}+\bar{\chi}_{(3,2)}\right) \\
    &\quad+ \left(\chi_{(3,1)} + \chi_{(3,4)}+ \omega_3^{2k}\,(\chi_{(1,4)}+\chi_{(4,1)} )+ \omega_3^k\,(\chi_{(1,3)} +\chi_{(4,3)})\right)\left(\bar{\chi}_{(1,3)}+\bar{\chi}_{(4,3)}\right)\,,\\

    Z_{\mathrm{PF}_1,\,\CE_8}[2,k] &= \left(\chi_{(1,6)}+\chi_{(3,2)}+\omega_3^{2k}\,(\chi_{(2,3)} +\chi_{(6,1)})+\omega_3^k\,(\chi_{(1,1)} +  \chi_{(3,3)} )\right)\left(\bar{\chi}_{(2,3)}+\bar{\chi}_{(6,1)}\right) \\
    &\quad+ \left(\chi_{(1,3)} +  \chi_{(4,3)}+\omega_3^{2k}\,(\chi_{(3,1)} + \chi_{(3,4)})+\omega_3^k\,(\chi_{(1,4)}+\chi_{(4,1)} )\right)\left(\bar{\chi}_{(3,1)}+\bar{\chi}_{(3,4)}\right)\,.
\end{aligned}
\end{align}
The other parafermionization with parameter $\rho=2$ yields another theory, whose partition functions are 
\begin{align}
\begin{aligned}
\label{eq:parafe8_2}
    Z_{\mathrm{PF_2},\,\CE_8}[0,k] &= \left({\chi}_{(1,1)}+{\chi}_{(3,3)}\right)\left(\bar{\chi}_{(1,1)} +\bar{\chi}_{(3,3)}+\omega_3^{2k}\, (\bar{\chi}_{(1,6)}+\bar{\chi}_{(3,2)})+\omega_3^k\,(\bar{\chi}_{(2,3)}+\bar{\chi}_{(6,1)})\right) \\
    &\quad+ \left({\chi}_{(1,4)}+{\chi}_{(4,1)}\right)\left(\bar{\chi}_{(1,4)}+\bar{\chi}_{(4,1)}+\omega_3^{2k}\,(\bar{\chi}_{(1,3)} + \bar{\chi}_{(4,3)})+\omega_3^k\,(\bar{\chi}_{(3,1)} + \bar{\chi}_{(3,4)})\right)\,,\\

    Z_{\mathrm{PF_2},\,\CE_8}[1,k] &= \left({\chi}_{(6,1)}+{\chi}_{(2,3)}\right)\left(\bar{\chi}_{(3,2)}+\bar{\chi}_{(1,6)} +\omega_3^{2k}\,(\bar{\chi}_{(2,3)}+\bar{\chi}_{(6,1)})+\omega_3^k\,(\bar{\chi}_{(1,1)}+ \bar{\chi}_{(3,3)})\right) \\
    &\quad+ \left({\chi}_{(3,1)}+{\chi}_{(3,4)}\right)\left(\bar{\chi}_{(1,3)} + \bar{\chi}_{(4,3)}+ \omega_3^{2k}\,(\bar{\chi}_{(1,3)}+\bar{\chi}_{(4,3)} )+ \omega_3^k\,(\bar{\chi}_{(1,4)} +\bar{\chi}_{(4,1)})\right)\,,\\

    Z_{\mathrm{PF}_2,\,\CE_8}[2,k] &= \left({\chi}_{(3,2)}+{\chi}_{(1,6)}\right)\left(\bar{\chi}_{(2,3)}+\bar{\chi}_{(6,1)}+\omega_3^{2k}\,(\chi_{(1,1)} +\chi_{(3,3)})+\omega_3^k\,(\chi_{(1,6)} +  \chi_{(3,2)} )\right) \\
    &\quad+ \left({\chi}_{(1,3)}+{\chi}_{(4,3)}\right)\left(\bar{\chi}_{(3,1)} +  \chi_{(3,4)}+\omega_3^{2k}\,(\chi_{(1,4)} + \chi_{(4,1)})+\omega_3^k\,(\chi_{(1,3)}+\chi_{(4,3)} )\right)\,.
\end{aligned}
\end{align}
Similarly, we can start with the charge-conjugated partition function $\CE_8^\ast$.
Since the twisted partition functions of $\CE_8^\ast$ are the charge conjugate of \eqref{eq:twste8}, its parafermionic partition functions are also the charge conjugate of \eqref{eq:parafe8} for $\rho=1$ and \eqref{eq:parafe8_2} for $\rho=2$.
While the untwisted partition function $Z_{\mathrm{PF}_{1,2},\,\CE_8}[0,0]$ is invariant under charge conjugation, the other twisted partition functions $Z_{\mathrm{PF}_{1,2},\,\CE_8}[a,b]$ ($a\neq0, b\neq0$) are not self-conjugate.
Thus, the two parafermionic theories are distinct: $\mathrm{PF}[\CE_8]\not= \mathrm{PF}[\CE_8^\ast]$ in the dependence on a paraspin structure.
We show the relationship of parafermionization associated with the modular invariants $\CE_8$ and $\CE_8^\ast$ in the left of Fig.~\ref{fig:para-ce8}.

\begin{figure}
    \centering
    \begin{minipage}{0.48\textwidth}
    \begin{tikzpicture}[transform shape, scale=0.9]
    \draw (0,0)node[font=\large]{$\CE_8^\ast$};
    \draw[<->,>=stealth] (3.75,-0.0) to node[below,font=\small]{bosonize} node[above,font=\small]{parafermionize}(0.75,-0.0);
    \draw (4.5,0)node[font=\large]{$\mathrm{PF}[\CE_8^\ast]$};
    \draw (0,3)node[font=\large]{$\CE_8$};
    \draw[<->,>=stealth] (0.75,3) to node[below,font=\small]{bosonize} node[above,font=\small]{parafermionize} (3.75,3.0);
    \draw (4.5,3) node[font=\large]{$\mathrm{PF}[\CE_8]$};
    \draw[<->,>=stealth] (4.5,0.5) to node[right,font=\small,align=left]{stack\\+conjugate} (4.5,2.5);
    \draw[<->,>=stealth] (0,0.5) to node[left,font=\small]{orbifold} (0,2.5);
    \end{tikzpicture}
    \end{minipage}\qquad
    \begin{minipage}{0.4\textwidth}
    \begin{tikzpicture}[transform shape, scale=0.9]
    \draw (0,0)node[font=\large]{$\CE_{12}$};
    \draw[<->,>=stealth] (3.5,-0.0) to node[below,font=\small]{bosonize} node[above,font=\small]{parafermionize}(0.5,-0.0);
    \draw (4.2,0)node[font=\large]{$\mathrm{PF}[\CE_{12}]$};
    \draw[->,>=stealth] (-0.4,0.3) arc (225:-45:0.45);
    \draw[->,>=stealth] (3.8,0.3) arc (225:-45:0.45);
    \draw (-0.2,1.4) node[font=\small] {orbifold};
    \draw (4,1.1) node[above,font=\small,align=center] {stack\\+conjugate};
    \end{tikzpicture}
    \end{minipage}
    \caption{The parafermionization of modular invariants $\CE_8^{(\ast)}$ in the left and $\CE_{12}$ in the right. We denote the parafermionization of the theory $\CT$ by $\mathrm{PF}[\CT]$.}
    \label{fig:para-ce8}
\end{figure}

\subsection{Parafermionic models from \texorpdfstring{$\CE_{12}$}{E12}}

Consider the exceptional modular invariant $\CE_{12}=\CE_{12}^{(\ast)}$ at level $k=9$.
While the other exceptional modular invariants $\CE_{12}'^{(\ast)}$ at the same level have no symmetry, the modular invariant $\CE_{12}=\CE_{12}^{(\ast)}$ has a non-anomalous $\BZ_3$ symmetry. 
Thus, we can parafermionize this theory using the $\BZ_3$ symmetry.
The twisted partition functions are
\begin{align}
    \begin{aligned}
        Z_{\CE_{12}}[0,k] &= |\chi_\mathbf{1}|^2 + (\omega_3^k+\omega_3^{2k}) \,|\chi_\mathbf{3}|^2\,,\\
        Z_{\CE_{12}}[1,k] &= |\chi_\mathbf{3}|^2 + \omega_3^k\,\chi_\mathbf{1}\,\bar{\chi}_\mathbf{3} + \omega_3^{2k}\,\chi_\mathbf{3}\,\bar{\chi}_\mathbf{1}\,,\\
        Z_{\CE_{12}}[2,k] &= |\chi_\mathbf{3}|^2 + \omega_3^{2k}\,\chi_\mathbf{1}\,\bar{\chi}_\mathbf{3} + \omega_3^{k}\,\chi_\mathbf{3}\,\bar{\chi}_\mathbf{1}\,.
    \end{aligned}
\end{align}
Here, to simplify the notation, we defined $\chi_\mathbf{1}=\chi_{(1,1)}+\chi_{(1,10)}+\chi_{(10,1)}+\chi_{(2,5)}+\chi_{(5,2)}+\chi_{(5,5)}$ and $\chi_\mathbf{3} = \chi_{(3,3)}+\chi_{(3,6)}+\chi_{(6,3)}$, which is based on the conformal embedding $\widehat{su}(3)_9\subset \widehat{e}(6)_1$.
There are three primaries $a,b,c$ in $\widehat{e}(6)_1$, whose conformal dimensions are $h_\mathrm{a}=0$ and $h_\mathrm{b}=h_\mathrm{c}=2/3$.
Also, their characters are given by $\chi_a =\chi_\mathbf{1}$ and $\chi_b =\chi_c =\chi_\mathbf{3}$~\cite{Christe:1988vc}.
Note that the new characters $\chi_\mathbf{1}$ and $\chi_\mathbf{3}$ are invariant under charge conjugation $C(a,b)=(b,a)$.
The parafermionized $\CE_{12}$ theory with $\rho=1$ has the partition functions
\begin{align}
    \begin{aligned}
    \label{eq:para_e12_1}
        Z_{\mathrm{PF}_1,\,\CE_{12}}[0,k] &= \left(\chi_{\mathbf{1}}+(\omega_3^k+\omega_3^{2k})\,\chi_{\mathbf{3}}\right) \bar{\chi}_\mathbf{1}\,,\\
        Z_{\mathrm{PF}_1,\,\CE_{12}}[1,k] &= \omega_3^{2k}\left(\chi_{\mathbf{1}}+(\omega_3^k+\omega_3^{2k})\,\chi_{\mathbf{3}}\right) \bar{\chi}_\mathbf{3}\,,\\
        Z_{\mathrm{PF}_1,\,\CE_{12}}[2,k] &= \omega_3^k\left(\chi_{\mathbf{1}}+(\omega_3^k+\omega_3^{2k})\,\chi_{\mathbf{3}}\right) \bar{\chi}_\mathbf{3}\,.
    \end{aligned}
\end{align}
These partition functions are invariant under the operation of stacking and conjugation, which is consistent with the self-duality of $\CE_{12}$ under orbifold. We show the relation between the modular invariant $\CE_{12}$ and the parafermionized theory in the right of Fig.~\ref{fig:para-ce8}.

One can also consider the other $\BZ_3$ parafermionization with $\rho=2$.
The torus partition functions $Z_{\mathrm{PF}_2,\,\CE_{12}}[a,b]$ are given by
\begin{align}
    \begin{aligned}
    \label{eq:para_e12_2}
        Z_{\mathrm{PF}_2,\,\CE_{12}}[0,k] &= \chi_\mathbf{1}\left(\bar{\chi}_{\mathbf{1}}+(\omega_3^k+\omega_3^{2k})\,\bar{\chi}_{\mathbf{3}}\right)\,,\\
        Z_{\mathrm{PF}_2,\,\CE_{12}}[1,k] &= \omega_3^{k}\,\chi_\mathbf{3}\left(\bar{\chi}_{\mathbf{1}}+(\omega_3^k+\omega_3^{2k})\,\bar{\chi}_{\mathbf{3}}\right) \,,\\
        Z_{\mathrm{PF}_2,\,\CE_{12}}[2,k] &= \omega_3^{2k}\,\chi_\mathbf{3}\left(\bar{\chi}_{\mathbf{1}}+(\omega_3^k+\omega_3^{2k})\,\bar{\chi}_{\mathbf{3}}\right) \,.
    \end{aligned}
\end{align}
The parafermionic theories~\eqref{eq:para_e12_1} and \eqref{eq:para_e12_2} are self-dual under the charge conjugation $C(a,b) = (b,a)$ since the new characters $\chi_\mathbf{1}$ and $\chi_\mathbf{3}$ are self-conjugate.

\section{Discussion}
\label{sec:discussion}

We have considered two-dimensional CFTs with the affine $\widehat{su}(2)$ and $\widehat{su}(3)$ algebra symmetries.
The bosonic modular-invariant partition functions have already been classified in the ADE classification.
We generalized the classification to the fermionic and parafermionic theories with the same affine symmetries.
As in the ADE classification, we found the relationship between (para)fermionic $\widehat{su}(2)$ theories and the non-simply Dynkin diagrams, which leads to the BCFG classification.

For $\widehat{su}(2)$ theories, we found the correspondence to non-simply Dynkin diagrams.
The relationship between the bosonic theories labeled by ADE and the (para)fermionic theories labeled by BCFG can be summarized in Fig.~\ref{fig:su2_ferm} and Fig.~\ref{fig:paraf_d4}.
This relation is similar to the folding of the simply laced Dynkin diagrams.
When a simply laced Dynkin diagram has an automorphism, then one can perform the folding and obtain a non-simply laced one:
\begin{equation}
    \text{Folding: }\quad A_{2\ell-1} \to C_\ell\,,\quad D_{\ell+1}\to B_\ell\,,\quad E_6\to F_4\,,\quad D_4\to G_2\,,
\end{equation}
where $\ell$ is a positive integer.
Note that one cannot fold $A_{2\ell}$ for $\ell\in\BZ$, while it has an $\BZ_2$ automorphism.
This corresponds to the non-anomalous condition on the CFT side.
Compared with Fig.~\ref{fig:su2_ferm} and Fig.~\ref{fig:paraf_d4}, we can see that the fermionization and parafermionization map can be understood as folding in terms of the corresponding Dynkin diagrams.
This extends the already-known correspondence between modular invariants and simply-laced Dynkin diagrams.
It would be interesting to pursue the origin of the extended correspondence between our $\widehat{su}(2)$ theories and Dynkin diagrams.

This work gives the classification of 2d fermionic and parafermionic CFTs with $\widehat{su}(2)$ and $\widehat{su}(3)$ symmetries.
In terms of 3d topological field theories, each torus partition function gives the corresponding topological interface~\cite{Fuchs:2002cm,Kapustin:2010if,Fuchs:2012dt}.
From the topological interfaces in 3d picture, the authors of~\cite{Gaiotto:2020iye} discuss the constraints on the renormalization group flow of bosonic $\widehat{su}(2)$ WZW models.
We may be able to generalize their discussion to fermionic and parafermionic cases since we have the list of topological interfaces.

Our models may serve as a testing ground for studying profiles of parafermionic CFTs.
Although a systematic approach to parafermionization has been established, the properties of parafermionic theories remain unclear, as only a limited number of parafermionic models have been constructed. 
Representative models include parafermionic minimal models~\cite{Yao:2020dqx}, parafermionized coset models~\cite{Duan:2023ykn}, and composite parafermionic models~\cite{Fukusumi:2022kly}.
It would be interesting to study topological line defects in (para)fermionic CFTs~\cite{Chang:2018iay,Chang:2022hud,Haghighat:2023sax,Chen:2023jht} as well as conformal boundary states (see~\cite{Smith:2021luc,Fukusumi:2021zme,Ebisu:2021acm} for fermionization).

\acknowledgments
We are grateful to K.~Ohmori and Y.~Fukusumi for their valuable discussions.
This work was supported by FoPM, WINGS Program, the University of Tokyo, and by JSPS KAKENHI Grant-in-Aid for JSPS fellows Grant No.\,23KJ0436.

\bibliographystyle{JHEP}
\bibliography{ref}

\providecommand{\href}[2]{#2}\begingroup\raggedright\begin{thebibliography}{10}

\bibitem{Cappelli:1986hf}
A.~Cappelli, C.~Itzykson, and J.~B. Zuber, {\it {Modular invariant partition functions in two dimensions}},  {\em Nucl. Phys. B} {\bf 280} (1987) 445--465.

\bibitem{Cappelli:1987xt}
A.~Cappelli, C.~Itzykson, and J.~B. Zuber, {\it {The ADE Classification of Minimal and A1(1) Conformal Invariant Theories}},  {\em Commun. Math. Phys.} {\bf 113} (1987) 1.

\bibitem{Kato:1987td}
A.~Kato, {\it {Classification of Modular Invariant Partition Functions in Two-dimensions}},  {\em Mod. Phys. Lett. A} {\bf 2} (1987) 585.

\bibitem{Gannon:1999cp}
T.~Gannon, {\it {The Cappelli-Itzykson-Zuber A-D-E classification}},  {\em Rev. Math. Phys.} {\bf 12} (2000) 739--748, [\href{http://arxiv.org/abs/math/9902064}{{\tt math/9902064}}].

\bibitem{Gannon:1992ty}
T.~Gannon, {\it {The Classification of affine SU(3) modular invariant partition functions}},  {\em Commun. Math. Phys.} {\bf 161} (1994) 233--264, [\href{http://arxiv.org/abs/hep-th/9212060}{{\tt hep-th/9212060}}].

\bibitem{Gannon:1994cf}
T.~Gannon, {\it {The Classification of SU(3) modular invariants revisited}},  {\em Ann. Inst. H. Poincare Phys. Theor.} {\bf 65} (1996) 15--56, [\href{http://arxiv.org/abs/hep-th/9404185}{{\tt hep-th/9404185}}].

\bibitem{Degiovanni:1989ne}
P.~Degiovanni, {\it {Z / NZ CONFORMAL FIELD THEORIES}},  {\em Commun. Math. Phys.} {\bf 127} (1990) 71.

\bibitem{Gannon:1992nq}
T.~Gannon, {\it {WZW commutants, lattices, and level 1 partition functions}},  {\em Nucl. Phys. B} {\bf 396} (1993) 708--736, [\href{http://arxiv.org/abs/hep-th/9209043}{{\tt hep-th/9209043}}].

\bibitem{Gannon:1992np}
T.~Gannon, {\it {Partition functions for heterotic WZW conformal field theories}},  {\em Nucl. Phys. B} {\bf 402} (1993) 729--753, [\href{http://arxiv.org/abs/hep-th/9209042}{{\tt hep-th/9209042}}].

\bibitem{DiFrancesco:1991st}
P.~Di~Francesco, {\it {Integrable lattice models, graphs and modular invariant conformal field theories}},  {\em Int. J. Mod. Phys. A} {\bf 7} (1992) 407--500.

\bibitem{Petkova:1995fw}
V.~B. Petkova and J.~B. Zuber, {\it {From CFT to graphs}},  {\em Nucl. Phys. B} {\bf 463} (1996) 161--193, [\href{http://arxiv.org/abs/hep-th/9510175}{{\tt hep-th/9510175}}].

\bibitem{Petkova:1996yv}
V.~B. Petkova and J.~B. Zuber, {\it {Conformal field theory and graphs}},  in {\em {21st International Colloquium on Group Theoretical Methods in Physics}}, 7, 1996.
\newblock \href{http://arxiv.org/abs/hep-th/9701103}{{\tt hep-th/9701103}}.

\bibitem{Zuber:2000ia}
J.-B. Zuber, {\it {CFT, BCFT, ADE and all that}},  in {\em {School: Bariloche 2000: Quantum Symmetries in Theoretical Physics and Mathematics}}, 1, 2000.
\newblock \href{http://arxiv.org/abs/hep-th/0006151}{{\tt hep-th/0006151}}.

\bibitem{Bae:2020xzl}
J.-B. Bae, Z.~Duan, K.~Lee, S.~Lee, and M.~Sarkis, {\it {Fermionic rational conformal field theories and modular linear differential equations}},  {\em PTEP} {\bf 2021} (2021), no.~8 08B104, [\href{http://arxiv.org/abs/2010.12392}{{\tt arXiv:2010.12392}}].

\bibitem{Bae:2021mej}
J.-B. Bae, Z.~Duan, K.~Lee, S.~Lee, and M.~Sarkis, {\it {Bootstrapping fermionic rational CFTs with three characters}},  {\em JHEP} {\bf 01} (2022) 089, [\href{http://arxiv.org/abs/2108.01647}{{\tt arXiv:2108.01647}}].

\bibitem{Duan:2022kxr}
Z.~Duan, K.~Lee, S.~Lee, and L.~Li, {\it {On classification of fermionic rational conformal field theories}},  {\em JHEP} {\bf 02} (2023) 079, [\href{http://arxiv.org/abs/2210.06805}{{\tt arXiv:2210.06805}}].

\bibitem{bruillard2017fermionic}
P.~Bruillard, C.~Galindo, T.~Hagge, S.-H. Ng, J.~Y. Plavnik, E.~C. Rowell, and Z.~Wang, {\it Fermionic modular categories and the 16-fold way},  {\em Journal of Mathematical Physics} {\bf 58} (2017), no.~4.

\bibitem{Cho:2022kzf}
G.~Y. Cho, H.-c. Kim, D.~Seo, and M.~You, {\it {Classification of fermionic topological orders from congruence representations}},  {\em Phys. Rev. B} {\bf 108} (2023), no.~11 115103, [\href{http://arxiv.org/abs/2210.03681}{{\tt arXiv:2210.03681}}].

\bibitem{Karch:2019lnn}
A.~Karch, D.~Tong, and C.~Turner, {\it {A Web of 2d Dualities: ${\bf Z}_2$ Gauge Fields and Arf Invariants}},  {\em SciPost Phys.} {\bf 7} (2019) 007, [\href{http://arxiv.org/abs/1902.05550}{{\tt arXiv:1902.05550}}].

\bibitem{Tachikawalec}
Y.~Tachikawa, ``{Topological phases and relativistic QFTs}.'' \url{https://member.ipmu.jp/yuji.tachikawa/lectures/2018-cern-rikkyo/}.
\newblock {notes of the lectures given in the CERN winter school, February 2018.}

\bibitem{Runkel:2020zgg}
I.~Runkel and G.~M.~T. Watts, {\it {Fermionic CFTs and classifying algebras}},  {\em JHEP} {\bf 06} (2020) 025, [\href{http://arxiv.org/abs/2001.05055}{{\tt arXiv:2001.05055}}].

\bibitem{Hsieh:2020uwb}
C.-T. Hsieh, Y.~Nakayama, and Y.~Tachikawa, {\it {Fermionic minimal models}},  {\em Phys. Rev. Lett.} {\bf 126} (2021), no.~19 195701, [\href{http://arxiv.org/abs/2002.12283}{{\tt arXiv:2002.12283}}].

\bibitem{Kulp:2020iet}
J.~Kulp, {\it {Two More Fermionic Minimal Models}},  {\em JHEP} {\bf 03} (2021) 124, [\href{http://arxiv.org/abs/2003.04278}{{\tt arXiv:2003.04278}}].

\bibitem{BoyleSmith:2023xkd}
P.~Boyle~Smith, Y.-H. Lin, Y.~Tachikawa, and Y.~Zheng, {\it {Classification of chiral fermionic CFTs of central charge $\le$ 16}},  {\em SciPost Phys.} {\bf 16} (2024), no.~2 058, [\href{http://arxiv.org/abs/2303.16917}{{\tt arXiv:2303.16917}}].

\bibitem{Rayhaun:2023pgc}
B.~C. Rayhaun, {\it {Bosonic rational conformal field theories in small genera, chiral fermionization, and symmetry/subalgebra duality}},  {\em J. Math. Phys.} {\bf 65} (2024), no.~5 052301, [\href{http://arxiv.org/abs/2303.16921}{{\tt arXiv:2303.16921}}].

\bibitem{Hohn:2023auw}
G.~H\"ohn and S.~M\"oller, {\it {Classification of Self-Dual Vertex Operator Superalgebras of Central Charge at Most 24}},  \href{http://arxiv.org/abs/2303.17190}{{\tt arXiv:2303.17190}}.

\bibitem{Yao:2020dqx}
Y.~Yao and A.~Furusaki, {\it {Parafermionization, bosonization, and critical parafermionic theories}},  {\em JHEP} {\bf 04} (2021) 285, [\href{http://arxiv.org/abs/2012.07529}{{\tt arXiv:2012.07529}}].

\bibitem{Lienart:2000jw}
S.~Lienart, P.~Ruelle, and O.~Verhoeven, {\it {On discrete symmetries in su(2) and su(3) affine theories and related graphs}},  {\em Nucl. Phys. B} {\bf 592} (2001) 479--511, [\href{http://arxiv.org/abs/hep-th/0007095}{{\tt hep-th/0007095}}].

\bibitem{Mukhi:2022bte}
S.~Mukhi and B.~C. Rayhaun, {\it {Classification of Unitary RCFTs with Two Primaries and Central Charge Less Than 25}},  {\em Commun. Math. Phys.} {\bf 401} (2023), no.~2 1899--1949, [\href{http://arxiv.org/abs/2208.05486}{{\tt arXiv:2208.05486}}].

\bibitem{Gowdigere:2023xnm}
C.~N. Gowdigere, S.~Kala, and J.~Santara, {\it {Classifying three-character RCFTs with Wronskian index equalling 3 or 4}},  \href{http://arxiv.org/abs/2308.01149}{{\tt arXiv:2308.01149}}.

\bibitem{Sugawara:1967rw}
H.~Sugawara, {\it {A Field theory of currents}},  {\em Phys. Rev.} {\bf 170} (1968) 1659--1662.

\bibitem{Christe:1988vc}
P.~Christe and F.~Ravanini, {\it {$G(N$) X $G$(l) / $G(N$)+l Conformal Field Theories and Their Modular Invariant Partition Functions}},  {\em Int. J. Mod. Phys. A} {\bf 4} (1989) 897.

\bibitem{Moore:1988ss}
G.~W. Moore and N.~Seiberg, {\it {Naturality in Conformal Field Theory}},  {\em Nucl. Phys. B} {\bf 313} (1989) 16--40.

\bibitem{DiFrancesco:1997nk}
P.~Di~Francesco, P.~Mathieu, and D.~Senechal, {\em {Conformal Field Theory}}.
\newblock Graduate Texts in Contemporary Physics. Springer-Verlag, New York, 1997.

\bibitem{Schultz:1964fv}
T.~D. Schultz, D.~C. Mattis, and E.~H. Lieb, {\it {Two-dimensional Ising model as a soluble problem of many fermions}},  {\em Rev. Mod. Phys.} {\bf 36} (1964) 856--871.

\bibitem{Fradkin:1980th}
E.~H. Fradkin and L.~P. Kadanoff, {\it {DISORDER VARIABLES AND PARAFERMIONS IN TWO-DIMENSIONAL STATISTICAL MECHANICS}},  {\em Nucl. Phys. B} {\bf 170} (1980) 1--15.

\bibitem{Polchinski:1998rq}
J.~Polchinski, {\em {String theory. Vol. 1: An introduction to the bosonic string}}.
\newblock Cambridge Monographs on Mathematical Physics. Cambridge University Press, 12, 2007.

\bibitem{Dixon:1985jw}
L.~J. Dixon, J.~A. Harvey, C.~Vafa, and E.~Witten, {\it {Strings on Orbifolds}},  {\em Nucl. Phys. B} {\bf 261} (1985) 678--686.

\bibitem{Dixon:1986jc}
L.~J. Dixon, J.~A. Harvey, C.~Vafa, and E.~Witten, {\it {Strings on Orbifolds. 2.}},  {\em Nucl. Phys. B} {\bf 274} (1986) 285--314.

\bibitem{Dixon:1986qv}
L.~J. Dixon, D.~Friedan, E.~J. Martinec, and S.~H. Shenker, {\it {The Conformal Field Theory of Orbifolds}},  {\em Nucl. Phys. B} {\bf 282} (1987) 13--73.

\bibitem{Vafa:1989ih}
C.~Vafa, {\it {Quantum Symmetries of String Vacua}},  {\em Mod. Phys. Lett. A} {\bf 4} (1989) 1615.

\bibitem{Bhardwaj:2017xup}
L.~Bhardwaj and Y.~Tachikawa, {\it {On finite symmetries and their gauging in two dimensions}},  {\em JHEP} {\bf 03} (2018) 189, [\href{http://arxiv.org/abs/1704.02330}{{\tt arXiv:1704.02330}}].

\bibitem{Numasawa:2017crf}
T.~Numasawa and S.~Yamaguchi, {\it {Mixed global anomalies and boundary conformal field theories}},  {\em JHEP} {\bf 11} (2018) 202, [\href{http://arxiv.org/abs/1712.09361}{{\tt arXiv:1712.09361}}].

\bibitem{Lin:2019kpn}
Y.-H. Lin and S.-H. Shao, {\it {Anomalies and Bounds on Charged Operators}},  {\em Phys. Rev. D} {\bf 100} (2019), no.~2 025013, [\href{http://arxiv.org/abs/1904.04833}{{\tt arXiv:1904.04833}}].

\bibitem{Lin:2021udi}
Y.-H. Lin and S.-H. Shao, {\it {$\mathbb{Z}_N$ symmetries, anomalies, and the modular bootstrap}},  {\em Phys. Rev. D} {\bf 103} (2021), no.~12 125001, [\href{http://arxiv.org/abs/2101.08343}{{\tt arXiv:2101.08343}}].

\bibitem{Ginsparg:1988ui}
P.~H. Ginsparg, {\it {APPLIED CONFORMAL FIELD THEORY}},  in {\em {Les Houches Summer School in Theoretical Physics: Fields, Strings, Critical Phenomena}}, 9, 1988.
\newblock \href{http://arxiv.org/abs/hep-th/9108028}{{\tt hep-th/9108028}}.

\bibitem{Fukusumi:2023psx}
Y.~Fukusumi, G.~Ji, and B.~Yang, {\it {Operator-state correspondence in simple current extended conformal field theories: Toward a general understanding of chiral conformal field theories and topological orders}},  \href{http://arxiv.org/abs/2311.15621}{{\tt arXiv:2311.15621}}.

\bibitem{Ji:2019ugf}
W.~Ji, S.-H. Shao, and X.-G. Wen, {\it {Topological Transition on the Conformal Manifold}},  {\em Phys. Rev. Res.} {\bf 2} (2020), no.~3 033317, [\href{http://arxiv.org/abs/1909.01425}{{\tt arXiv:1909.01425}}].

\bibitem{Bae:2021lvk}
J.-B. Bae and S.~Lee, {\it {Emergent supersymmetry on the edges}},  {\em SciPost Phys.} {\bf 11} (2021), no.~5 091, [\href{http://arxiv.org/abs/2105.02148}{{\tt arXiv:2105.02148}}].

\bibitem{Kikuchi:2022jbl}
K.~Kikuchi, {\it {Emergent SUSY in two dimensions}},  \href{http://arxiv.org/abs/2204.03247}{{\tt arXiv:2204.03247}}.

\bibitem{Fukusumi:2022ucr}
Y.~Fukusumi, {\it {Gaplessness protected by bulk-edge correspondence}},  \href{http://arxiv.org/abs/2212.12996}{{\tt arXiv:2212.12996}}.

\bibitem{Kitaev:2000nmw}
A.~Kitaev, {\it {Unpaired Majorana fermions in quantum wires}},  {\em Phys. Usp.} {\bf 44} (2001), no.~10S 131--136, [\href{http://arxiv.org/abs/cond-mat/0010440}{{\tt cond-mat/0010440}}].

\bibitem{Fateev:1987vh}
V.~A. Fateev and A.~B. Zamolodchikov, {\it {Conformal quantum field theory models in two dimensions having Z3 symmetry}},  {\em Nucl. Phys. B} {\bf 280} (1987) 644--660.

\bibitem{Gepner:1986hr}
D.~Gepner and Z.-a. Qiu, {\it {Modular Invariant Partition Functions for Parafermionic Field Theories}},  {\em Nucl. Phys. B} {\bf 285} (1987) 423.

\bibitem{Thorngren:2021yso}
R.~Thorngren and Y.~Wang, {\it {Fusion category symmetry. Part II. Categoriosities at c = 1 and beyond}},  {\em JHEP} {\bf 07} (2024) 051, [\href{http://arxiv.org/abs/2106.12577}{{\tt arXiv:2106.12577}}].

\bibitem{Burbano:2021loy}
I.~M. Burbano, J.~Kulp, and J.~Neuser, {\it {Duality defects in E$_{8}$}},  {\em JHEP} {\bf 10} (2022) 186, [\href{http://arxiv.org/abs/2112.14323}{{\tt arXiv:2112.14323}}].

\bibitem{Fukusumi:2022kly}
Y.~Fukusumi, {\it {Composing parafermions: a construction of $Z_{N}$ fractional quantum Hall systems and a modern understanding of confinement and duality}},  \href{http://arxiv.org/abs/2212.12999}{{\tt arXiv:2212.12999}}.

\bibitem{Duan:2023ykn}
Z.~Duan, Q.~Jia, and S.~Lee, {\it {\ensuremath{\mathbb{Z}}$_{N}$ duality and parafermions revisited}},  {\em JHEP} {\bf 11} (2023) 206, [\href{http://arxiv.org/abs/2309.01913}{{\tt arXiv:2309.01913}}].

\bibitem{alexandradinata2016parafermionic}
A.~Alexandradinata, N.~Regnault, C.~Fang, M.~J. Gilbert, and B.~A. Bernevig, {\it Parafermionic phases with symmetry breaking and topological order},  {\em Physical Review B} {\bf 94} (2016), no.~12 125103.

\bibitem{stekolshchik2005notes}
R.~Stekolshchik, {\it Notes on coxeter transformations and the mckay correspondence},  {\em arXiv preprint math/0510216} (2005).

\bibitem{Zuber:1993vm}
J.~B. Zuber, {\it {On Dubrovin topological field theories}},  {\em Mod. Phys. Lett. A} {\bf 9} (1994) 749--760, [\href{http://arxiv.org/abs/hep-th/9312209}{{\tt hep-th/9312209}}].

\bibitem{Schellekens:1986mb}
A.~N. Schellekens and N.~P. Warner, {\it {Conformal Subalgebras of {Kac-Moody} Algebras}},  {\em Phys. Rev. D} {\bf 34} (1986) 3092.

\bibitem{Bais:1986zs}
F.~A. Bais and P.~G. Bouwknegt, {\it {A Classification of Subgroup Truncations of the Bosonic String}},  {\em Nucl. Phys. B} {\bf 279} (1987) 561.

\bibitem{Fuchs:2002cm}
J.~Fuchs, I.~Runkel, and C.~Schweigert, {\it {TFT construction of RCFT correlators 1. Partition functions}},  {\em Nucl. Phys. B} {\bf 646} (2002) 353--497, [\href{http://arxiv.org/abs/hep-th/0204148}{{\tt hep-th/0204148}}].

\bibitem{Kapustin:2010if}
A.~Kapustin and N.~Saulina, {\it {Surface operators in 3d Topological Field Theory and 2d Rational Conformal Field Theory}},  \href{http://arxiv.org/abs/1012.0911}{{\tt arXiv:1012.0911}}.

\bibitem{Fuchs:2012dt}
J.~Fuchs, C.~Schweigert, and A.~Valentino, {\it {Bicategories for boundary conditions and for surface defects in 3-d TFT}},  {\em Commun. Math. Phys.} {\bf 321} (2013) 543--575, [\href{http://arxiv.org/abs/1203.4568}{{\tt arXiv:1203.4568}}].

\bibitem{Gaiotto:2020iye}
D.~Gaiotto and J.~Kulp, {\it {Orbifold groupoids}},  {\em JHEP} {\bf 02} (2021) 132, [\href{http://arxiv.org/abs/2008.05960}{{\tt arXiv:2008.05960}}].

\bibitem{Chang:2018iay}
C.-M. Chang, Y.-H. Lin, S.-H. Shao, Y.~Wang, and X.~Yin, {\it {Topological Defect Lines and Renormalization Group Flows in Two Dimensions}},  {\em JHEP} {\bf 01} (2019) 026, [\href{http://arxiv.org/abs/1802.04445}{{\tt arXiv:1802.04445}}].

\bibitem{Chang:2022hud}
C.-M. Chang, J.~Chen, and F.~Xu, {\it {Topological defect lines in two dimensional fermionic CFTs}},  {\em SciPost Phys.} {\bf 15} (2023), no.~5 216, [\href{http://arxiv.org/abs/2208.02757}{{\tt arXiv:2208.02757}}].

\bibitem{Haghighat:2023sax}
B.~Haghighat and Y.~Sun, {\it {Topological Defect Lines in bosonized Parafermionic CFTs}},  \href{http://arxiv.org/abs/2306.16555}{{\tt arXiv:2306.16555}}.

\bibitem{Chen:2023jht}
J.~Chen, B.~Haghighat, and Q.-R. Wang, {\it {Para-fusion Category and Topological Defect Lines in $\mathbb Z_N$-parafermionic CFTs}},  \href{http://arxiv.org/abs/2309.01914}{{\tt arXiv:2309.01914}}.

\bibitem{Smith:2021luc}
P.~B. Smith, {\it {Boundary States and Anomalous Symmetries of Fermionic Minimal Models}},  \href{http://arxiv.org/abs/2102.02203}{{\tt arXiv:2102.02203}}.

\bibitem{Fukusumi:2021zme}
Y.~Fukusumi, Y.~Tachikawa, and Y.~Zheng, {\it {Fermionization and boundary states in 1+1 dimensions}},  {\em SciPost Phys.} {\bf 11} (2021), no.~4 082, [\href{http://arxiv.org/abs/2103.00746}{{\tt arXiv:2103.00746}}].

\bibitem{Ebisu:2021acm}
H.~Ebisu and M.~Watanabe, {\it {Fermionization of conformal boundary states}},  {\em Phys. Rev. B} {\bf 104} (2021), no.~19 195124, [\href{http://arxiv.org/abs/2103.01101}{{\tt arXiv:2103.01101}}].

\end{thebibliography}\endgroup
\end{document}